\documentclass[reprint,amsmath,amssymb, notitlepage,prb]{revtex4-1}
\usepackage{graphicx}
\usepackage{bm}
\usepackage{mathrsfs}

\allowdisplaybreaks[1]
\usepackage{color}
\usepackage[normalem]{ulem}

\newcommand{\average}[1]{\ensuremath{\langle#1\rangle} }

\newcommand{\blue}[1]{{\color{blue}#1}}

\begin{document}
	
	\title{Spin Pumping into Anisotropic Dirac Electrons}
	\author{Takumi Funato${}^{1,2}$, Takeo Kato${}^3$, Mamoru Matsuo${}^{2,4,5,6}$}
	\affiliation{$^1$Center for Spintronics Research Network, Keio University, Yokohama 223-8522, Japan}
	\affiliation{$^2$
		Kavli Institute for Theoretical Sciences, University of Chinese Academy of Sciences, Beijing, 100190, China.
	}
    \affiliation{$^3$Institute for Solid State Physics, The University of Tokyo, Kashiwa, Japan}
	\affiliation{$^4$
		CAS Center for Excellence in Topological Quantum Computation, University of Chinese Academy of Sciences, Beijing 100190, China
	}%
	\affiliation{$^5$
		Advanced Science Research Center, Japan Atomic Energy Agency, Tokai, 319-1195, Japan
	}
	\affiliation{$^6$RIKEN Center for Emergent Matter Science (CEMS), Wako, Saitama 351-0198, Japan}

	\date{\today}
	
	\begin{abstract}
	We study spin pumping into an anisotropic Dirac electron system induced by microwave irradiation to an adjacent ferromagnetic insulator theoretically. 
    We formulate the Gilbert damping enhancement due to the spin current flowing into the Dirac electron system using second-order perturbation with respect to the interfacial exchange coupling. 
    As an illustration, we consider the anisotropic Dirac system realized in bismuth to show that the Gilbert damping varies according to the  magnetization direction in the ferromagnetic insulator.
    Our results indicate that this setup can provide helpful information on the anisotropy of the Dirac electron system.
	\end{abstract}

	\pacs{Valid PACS appear here}
	\maketitle
	
	\section{Introduction}
	In spintronics, spin currents are crucial in using electrons' charge and spin. 
	Spin pumping, the spin current generation of conduction electrons from nonequilibrium magnetization dynamics at magnetic interfaces, is a popular method for generating and manipulating spin currents. 
	In previous experimental reports on spin pumping, the enhancement of Gilbert damping in ferromagnetic resonance (FMR) was observed due to the loss of angular momentum associated with the spin current injection into the nonmagnetic layer adjacent to the ferromagnetic layer~\cite{heinrichFerromagneticresonanceStudyUltrathin1987,celinskiFerromagneticResonanceLinewidth1991,mizukamiFerromagneticResonanceLinewidth2001,mizukamiStudyFerromagneticResonance2001,mizukamiEffectSpinDiffusion2002,ingvarssonRoleElectronScattering2002,lubitzIncreaseMagneticDamping2003,sarmaSpintronicsFundamentalsApplications2004,tserkovnyakNonlocalMagnetizationDynamics2005}.
	Mizukami {\it et al.} measured the enhancement of the Gilbert damping associated with the adjacent nonmagnetic metal.
	They reported that the strong spin-orbit coupling in the nonmagnetic layer strictly affected the enhancement of the Gilbert damping~\cite{mizukamiFerromagneticResonanceLinewidth2001,mizukamiStudyFerromagneticResonance2001,mizukamiEffectSpinDiffusion2002}.
	Consequently, electric detection by inverse spin Hall effect, in which the charge current is converted from the spin current, led to spin pumping being used as an essential technique for studying spin-related phenomena in nonmagnetic materials~\cite{azevedoDcEffectFerromagnetic2005,saitohConversionSpinCurrent2006,andoElectricManipulationSpin2008,andoElectricDetectionSpin2009,andoElectricallySpin2011,mosendzDetectionQuantificationInverse2010,mosendzQuantifyingSpinHall2010,czeschkaScalingBehaviorSpin2011,mironPerpendicularSwitchingSingle2011,liuSpinTorqueFerromagneticResonance2011,liuSpinTorqueSwitchingGiant2012,kajiwaraTransmissionElectricalSignals2010,baiUniversalMethodSeparating2013,sandwegSpinPumpingParametrically2011,sinovaSpinHallEffects2015}.
	Saitoh {\it et al.} measured electric voltage in a bilayer of Py and Pt under microwave application.
	They observed that charge current converted because of inverse spin Hall effect from spin current injected by spin pumping~\cite{saitohConversionSpinCurrent2006}.
	
	In the first theoretical report on spin pumping, Berger predicted an increase in Gilbert damping due to the spin current flowing interface between the ferromagnetic and nonmagnetic layers~\cite{bergerEmissionSpinWaves1996,hellmanInterfaceinducedPhenomenaMagnetism2017}.
	Tserkovnyak {\it et al.} calculated the spin current flowing through the interface~\cite{tserkovnyakSpinPumpingMagnetization2002,tserkovnyakEnhancedGilbertDamping2002,tserkovnyakDynamicExchangeCoupling2003} based on the scattering-matrix theory and the picture of adiabatic spin pumping~\cite{muccioloAdiabaticQuantumPump2002,sharmaAdiabaticChargeSpin2003,watsonExperimentalRealizationQuantum2003}.
	They introduced a complex spin-mixing conductance that characterizes spin transport at the interfaces based on spin conservation and no spin loss.
	The spin mixing conductance can represent the spin pumping-associated phenomena and is quantitatively evaluated using the first principle calculation~\cite{xiaSpinTorquesFerromagnetic2002}.
	Nevertheless, microscopic analysis is necessary to understand the detailed mechanism of spin transport at the interface~\cite{ohnumaEnhanced2014,matsuo2018Spin,katoMicroscopicTheorySpin2019,katoMicroscopicTheorySpin2020,ominatoQuantumOscillationsGilbert2020,ominatoValleyDependentSpinTransport2020,yamamotoSpinCurrentMagnetic2021,ominatoAnisotropicSuperconductingSpin2021,yama2021Spin,Yama2022,ominato2022Ferromagnetica}.
	It was clarified that spin pumping depends on the anisotropy of the electron band structure and spin texture. Spin pumping is expected to be one of the probes of the electron states~\cite{ominatoAnisotropicSuperconductingSpin2021,yama2021Spin,Yama2022,ominato2022Ferromagnetica}.

	\begin{figure}[b]
	    \centering
	    \includegraphics[width=75mm]{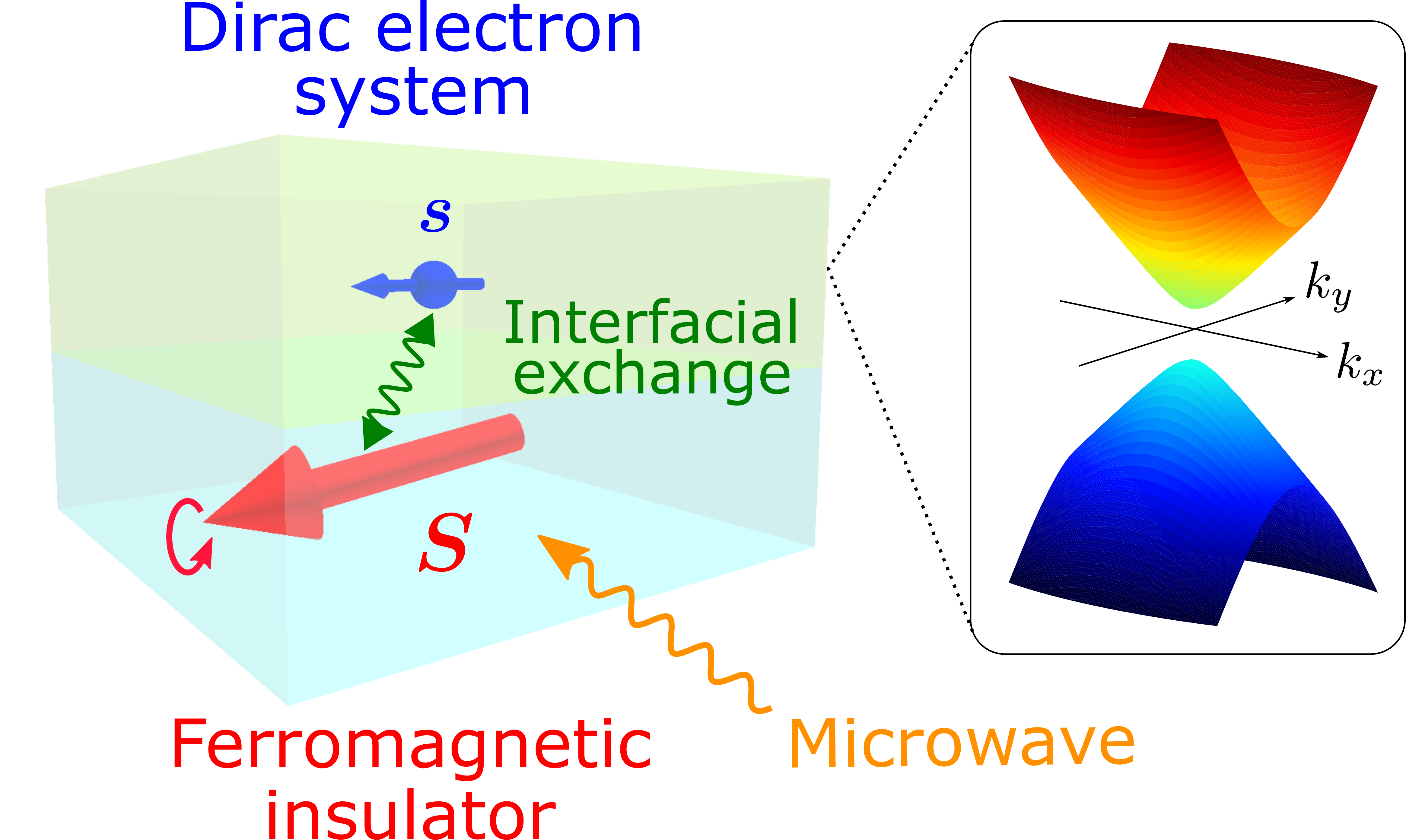}
	    \caption{
	    Schematic illustration of a bilayer system composed of the Dirac electron system and ferromagnetic insulator.
	    The applied microwave excited precession of the localized spin in the ferromagnetic insulator and spin current is injected into the Dirac electron system.
	    }
	    \label{fig1}
	\end{figure}
	
    Bismuth has been extensively studied because of its attractive physical properties, such as large diamagnetism, large $g$-factor, high efficient Seebeck effect, Subrikov-de Haas effect, and de Haas-van Alphen effect~\cite{edelmanElectronsBismuth1976,fuseyaTransportPropertiesDiamagnetism2015}.
    The electrons in the conduction and valence bands near the $L$-point in bismuth, which contribute mainly to the various physical phenomena, are expressed as effective Dirac electrons. 
    Thus, electrons in bismuth are called Dirac electrons~\cite{wolffMatrixElementsSelection1964,edelmanElectronsBismuth1976,fuseyaTransportPropertiesDiamagnetism2015}.
    The doping antimony to bismuth is known to close the gap and makes it a topological insulator~\cite{fuTopologicalInsulatorsInversion2007,teoSurfaceStatesTopological2008}.
    Because of its strong spin-orbit interaction, 
    bismuth has attracted broad attention in spintronics as a high efficient charge-to-spin conversion material~\cite{fuseyaSpinHallEffectDiamagnetism2012,fuseyaSpinPolarizationMagnetoOpticalConductivity2012,fuseyaSpinHallEffectDiamagnetism2014,fukazawaIntrinsicExtrinsicSpin2017,yueSpintoChargeConversionBi2018,chiSpinHallEffect2020}.
    The spin current generation at the interface between the bismuth oxide and metal has been studied since a significant Rashba spin-orbit interaction appears at the interface~\cite{karubeExperimentalObservationSpintocharge2016}.
    The spin injection into bismuth was observed due to spin pumping from yttrium iron garnet or permalloy~\cite{houInterfaceInducedInverse2012,emotoConversionPureSpin2014,emotoTransportSpinConversion2016}.
    Nevertheless, microscopic analysis of spin pumping into bismuth has not been performed.
    The dependence of the spin pumping on the crystal and band structure of bismuth remains unclear.

	This study aims at a microscopic analysis of spin injection due to spin pumping into an anisotropic Dirac electron system, such as bismuth,  and investigates the dependence of spin pumping on the band structure. 
	We consider a bilayer system comprising an anisotropic Dirac electron system and a ferromagnetic insulator where a microwave is applied (see Fig.~\ref{fig1}). 
	The effect of the interface is treated by proximity exchange coupling between the Dirac electron spins and the localized spins of the ferromagnetic insulator~\cite{ohnumaEnhanced2014,matsuo2018Spin,katoMicroscopicTheorySpin2019,katoMicroscopicTheorySpin2020,ominatoQuantumOscillationsGilbert2020,ominatoValleyDependentSpinTransport2020,yamamotoSpinCurrentMagnetic2021,ominatoAnisotropicSuperconductingSpin2021,yama2021Spin,Yama2022,ominato2022Ferromagnetica}.
	We calculate the Gilbert damping enhancement due to spin pumping from the ferromagnetic insulator into the Dirac electron system up to the second perturbation of the interfacial exchange coupling. 
	For illustarion, we calculate the enhancement of the Gilbert damping for an anisotropic Dirac system in bismuth.

	This paper is organized as follows:
	Sec.~\ref{sec.model} describes the model.
	Sec.~\ref{sec.formulation} shows the formulation of the Gilbert damping enhancement and discuss the effect of the interfacial randomness on spin pumping.
	Sec.~\ref{sec.result} summarizes the results and demonstration of the Gilbert damping enhancement in bismuth.
	Sec.~\ref{sec.conclusion} presents the conclusion.
	The Appendices show the details of the calculation.
	Appendix~\ref{spin_op} defines the magnetic moment of electrons in a Dirac electron system.
	Appendix~\ref{app:linearresponse} provides the detailed formulation of the Gilbert damping modulation, and Appendix~\ref{app:detailedderivation} presents the detailed derivation of Gilbert damping modulation.

	\section{Model\label{sec.model}}
	We consider a bilayer system composed of an anisotropic Dirac electron system and a ferromagnetic insulator under a static magnetic field.
    We evaluate a microscopic model whose Hamiltonian is given as
	\begin{align}
	    \hat{ \mathcal H}_T = \hat{ \mathcal H}_D + \hat {\mathcal H}_{\text{FI}} + \hat{\mathcal H}_{\text{ex}},
	    \label{total}
	\end{align}
	where $\hat{\mathcal H}_D$, $\hat{\mathcal H}_{\text{FI}}$, and $\hat{\mathcal H}_{\text{ex}}$ represent an anisotropic Dirac electron system, a ferromagnetic insulator, and an interfacial exchange interaction, respectively.
	
	\subsection{Anisotropic Dirac system}
	
	The following Wolff Hamiltonian models the anisotropic Dirac electron system~\cite{wolffMatrixElementsSelection1964,fuseyaSpinHallEffectDiamagnetism2012,fuseyaTransportPropertiesDiamagnetism2015}:
	\begin{align}
		\hat{\mathcal H}_D 
		=\sum_{\bm k}
		c^{\dagger}_{\bm k}(- \hbar \bm k\cdot \bm v \rho_2 + \Delta \rho_3) c_{\bm k},
	\end{align}
	where $2\Delta$ ($\neq 0$) is the band gap,  $c^{\dagger}_{\bm k}$($c_{\bm k}$) is the electrons' four-component creation (annihilation) operator, and $\bm v$ is the velocity operator given by $v_i = \sum_{\alpha} w_{i\alpha} \sigma^{\alpha}$ with $w_{i\alpha}$ being the matrix element of the velocity operator.
	$\bm \sigma =(\sigma^x,\sigma^y,\sigma^z)$ are the Pauli matrices in the spin space and $\bm \rho =(\rho_1,\rho_2,\rho_3)$ are the Pauli matrices specifying the conduction and valence bands.
		
	For this anisotropic Dirac system, the Matsubara Green function of the electrons is given by 
	\begin{align}
	    g_{\bm k}(i\epsilon_n) 
	    &=\frac{i\epsilon_n+\mu - \hbar \tilde{\bm k}\cdot \bm \sigma \rho_2 + \Delta \rho_3}{(i\epsilon_n+\mu)^2-\epsilon_{\bm k}^2},
	\end{align}
	where $\epsilon_n=(2n+1)\pi/\beta$ is the fermionic Matsubara frequencies with $n$ being integers, $\mu$ $(>\Delta)$ is the chemical potential in the conduction band $\tilde{\bm k}$ is defined by $\tilde{\bm k}\cdot \bm \sigma =\tilde k_{\alpha}\sigma^{\alpha}=\bm k\cdot \bm v$, and $\epsilon_{\bm k}$ is the eigenenergy given by
	\begin{align}
	    \epsilon_{\bm k} = \sqrt{\Delta^2 + (\hbar k_i w_{i\alpha})^2} = \sqrt{\Delta^2 + \hbar^2 \tilde k^2}.
	\end{align}
	
	The density of state of the Dirac electrons per unit cell per band and spin is givcen by
	\begin{align}
	    \nu(\epsilon) &= n_{\text D}^{-1} \sum_{\bm k,\lambda} \delta(\epsilon - \lambda \epsilon_{\bm k}),
	    \\
	    &= \frac{|\epsilon|}{2\pi^2 \hbar^3} \sqrt{ \frac{\epsilon^2-\Delta^2}{\Delta^3 \text{det}\alpha_{ij}}}  \theta(|\epsilon|-\Delta),
	\end{align}
	where $n_D$ is the number of unit cells in the system and $\alpha_{ij}$ is the inverse mass tensor near the bottom of the band, which characterize the band structure of the anisotropic Dirac electron system:
	\begin{align}
	    \alpha_{ij} =\left.\frac{1}{\hbar^2}\frac{\partial^2\epsilon_{\bm k}}{\partial k_i\partial k_j}
	    \right|_{\bm k=\bm 0}
	    = \frac{1}{\Delta}\sum_{\alpha}w_{i\alpha}w_{j\alpha}.
	\end{align}	
	
	The spin operator can be defined as
	\begin{align}
	    \hat{\bm s}_{\bm q} &= \sum_{\bm k} c^{\dagger}_{\bm k-\bm q/2} \bm s c_{\bm k+\bm q/2}, \\
	    s^i &= \frac{m}{\Delta }  \mathcal M_{i\alpha} \rho_3 \sigma^{\alpha}, \ \ \ \ (i=x,y,z),
	\end{align}
	where $M_{i\alpha}$ are the matrix elements of the spin magnetic moment given as~\cite{fuseyaSpinPolarizationMagnetoOpticalConductivity2012,fuseyaSpinHallEffectDiamagnetism2012}
	\begin{align}
	    \mathcal M_{i\alpha} = \epsilon_{\alpha \beta \gamma} \epsilon_{ijk} w_{i\beta}w_{j\gamma}/2.
	\end{align}
	The detailed derivation of the spin magnetic moment can be found in Appendix~\ref{spin_op}.

    \subsection{Ferromagnetic insulator}
    
    \begin{figure}
        \centering
        \includegraphics[width=50mm]{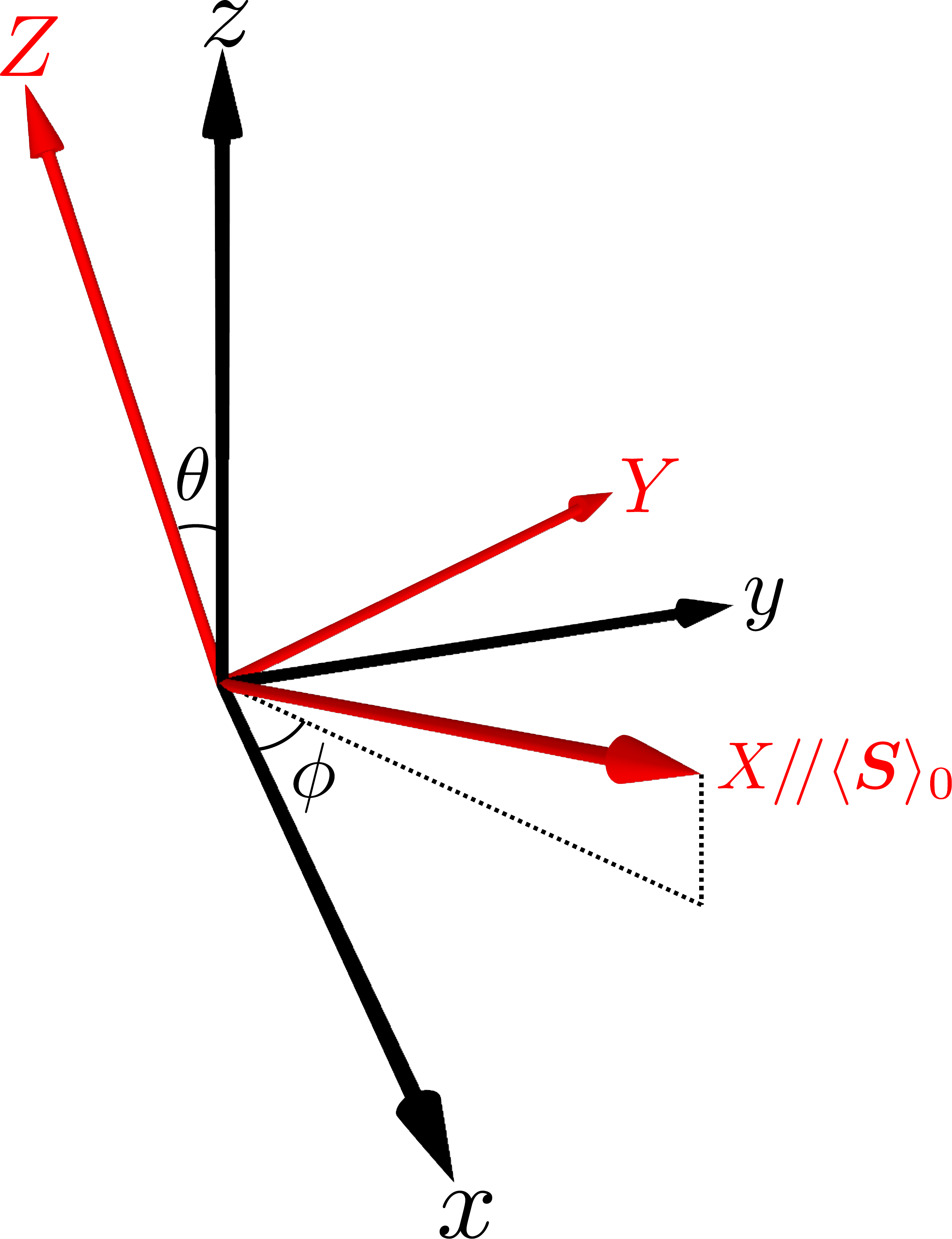}
        \caption{Relation between the original coordinates $(x,y,z)$ and the magnetization-fixed coordinates $(X,Y,Z)$.
        The direction of the ordered localized spin $\average{\bm S}_0$ is fixed to the $X$-axis.
        $\theta$ is the polar angle and $\phi$ is the azimuthal angle.
        }
        \label{coordinate}
    \end{figure}
    
    The bulk ferromagnetic insulator under a static magnetic field is described by the quantum Heisenberg model as
	\begin{align}
	    \hat{\mathcal H}_{\text{FI}} = -2J\sum_{\average{i,j}}\bm S_i \cdot \bm S_j - g\mu_{\text B} h_{\text{dc}} \sum_i S^X_i,
	    \label{app_hfi}
	\end{align}
	where $J$ is an exchange interaction, $g$ is g-factor of the electrons, $\mu_{\text B}$ is the Bohr magnetization, and $\average{i,j}$ represents the pair of nearest neighbor sites.
	Here, we have introduced a magnetization-fixed coordinate $(X,Y,Z)$, for which the direction of the ordered localized spin $\average{\bm S}_0$ is fixed to the $X$-axis.
	The localized spin operators for the magnetization-fixed coordinates are related to the ones for the original coordinates $(x,y,z)$ as
	\begin{gather}
	    \begin{pmatrix}
	        S^x\\S^y\\S^z
	    \end{pmatrix}
	    =\mathcal R (\theta,\phi) 
	     \begin{pmatrix}
	        S^X\\S^Y\\S^Z
	    \end{pmatrix},
	\end{gather}
	where $\mathcal R(\theta,\phi)= \mathcal R_z(\phi) \mathcal R_y(\theta)$ is the rotation matrix combining the polar angle $\theta$ rotation around the $y$-axis $\mathcal R_y(\theta)$ and the azimuthal angle $\phi$ rotation around the $z$-axis $\mathcal R_z(\phi)$, given by
	\begin{align}
	    \mathcal R (\theta,\phi)
	    =
	    \begin{pmatrix}
	    \cos \theta \cos \phi & -\sin \phi & \sin \theta \cos \phi \\
	    \cos \theta \sin \phi & \cos \phi & \sin \theta \sin \phi \\
	    -\sin \theta & 0 & \cos \theta 
	    \end{pmatrix}
	    .
	\end{align}	
	By applying the spin-wave approximation, the spin operators are written as $S^\pm_{\bm k}=S^Y_{\bm k}\pm iS^Z_{\bm k}= \sqrt{2S}b_{\bm k}(b_{\bm k}^{\dagger})$ and $S^X_{\bm k}=S-b^{\dagger}_{\bm k}b_{\bm k}$ using magnon creation/annihilation operators, $b_{\bm k}^{\dagger}$ and $b_{\bm k}$. Then, the Hamiltonian is rewritten as
	\begin{gather}
	    \hat{\mathcal H}_{\text{FI}} = \sum_{\bm k}\hbar \omega_{\bm k} b^{\dagger}_{\bm k} b_{\bm k},
	\end{gather}
	where $\hbar \omega _{\bm k}=\mathscr Dk^2+\hbar \omega_{\bm 0}$ with $\mathscr D = zJSa^2$ being the spin stiffness and $z$ being the number of the nearest neighbor sites, and $\hbar \omega_{\bm 0}=g\mu_{\text B}h_{\text{dc}}$ is the Zeeman energy.

	\subsection{Interfacial exchange interaction}
	
	The proximity exchange coupling between the electron spin in the anisotropic Dirac system and the localized spin in the ferromagnetic insulator is modeled by
	\begin{align}
		\hat{\mathcal H}_{\text{ex}} = \sum_{\bm q,\bm k} (\mathcal T_{\bm q,\bm k} \hat s^+_{\bm q}  S^-_{\bm k} + \text{h.c.} ),
	\end{align}
	where $\mathcal T_{\bm q,\bm k}$ is a matrix element for spin transfer through the interface and $\hat s^{\pm}_{\bm q}=\hat s^Y_{\bm q}\pm i\hat s^Z_{\bm q}$ are the spin ladder operators of the Dirac electrons.
	According to the relation between the original coordinate $(x,y,z)$ and the magnetization-fixed coordinate $(X,Y,Z)$, the spin operators of the Dirac electrons are expressed as
	\begin{align}
	    \begin{pmatrix}
	    s^X \\ s^Y \\ s^Z
	    \end{pmatrix}
	    = \mathcal R^{-1}(\theta,\phi)
	    \begin{pmatrix}
	    s^x \\ s^y \\ s^z
	    \end{pmatrix},
	\end{align}
	where $\mathcal R^{-1}(\theta,\phi)=\mathcal R_y(\theta)\mathcal R_z(-\phi)$ is given by
	\begin{align}
	    \mathcal R^{-1}(\theta,\phi) =
	    \begin{pmatrix}
	    \cos \theta \cos \phi & \cos \theta \sin \phi & -\sin \theta \\
	    -\sin \phi & \cos \phi & 0 \\
	    \sin \theta \cos \phi & \sin\theta \sin \phi & \cos \theta 
	    \end{pmatrix}
	    .
	\end{align}
	The spin ladder operators are given by
	\begin{gather}
	    s^+ = \frac{m}{\Delta } a_i \mathcal M_{i\alpha} \sigma^{\alpha},
	    \ \ \ \ 
	    s^- = \frac{m}{\Delta } a_i^* \mathcal M_{i\alpha} \sigma^{\alpha},
	\end{gather}
	where 
	$a_i$ ($i=x,y,z$) are defined by
	\begin{gather}
	    \begin{pmatrix}
	        a_x \\ a_y \\ a_z
	    \end{pmatrix}
	    =
	    \begin{pmatrix}
	        -\sin \phi + i\sin \theta \cos \phi\\
	    \cos \phi + i\sin \theta \sin \phi\\
	     i\cos \theta 
	    \end{pmatrix}
	    .
	\end{gather}
	
	\section{Formulation\label{sec.formulation}}
	
	Applying a microwave to the ferromagnetic insulator includes the localized spin's precession.
	The Gilbert damping constant can be read from the retarded magnon Green function defined by
	\begin{align}
	    G_{\bm k}^R(\omega) = -\frac{i}{\hbar}\int^{\infty}_0 dt e^{i(\omega+i\delta)t} \average{[S^+_{\bm k}(t),S^-_{\bm k}]},
	    \label{eq:GRomegadef}
	\end{align}
	with $S^+_{\bm k}(t)=e^{i\hat{ \mathcal H}_T/\hbar}S^+_{\bm k} e^{-i\hat{ \mathcal H}_T/\hbar}$ being the Heisenberg representation of the localized spin, since one can prove that the absorption rate of the microwave is proportional to ${\rm Im} \, G_{{\bm k}={\bm 0}}^R(\omega)$
	(see also Appendix~\ref{app:linearresponse}).
	By considering the second-order perturbation with respect to the matrix element for the spin transfer $\mathcal T_{\bm q,\bm k}$, the magnon Green function is given by~\cite{ohnumaEnhanced2014,matsuo2018Spin,katoMicroscopicTheorySpin2019,katoMicroscopicTheorySpin2020,ominatoQuantumOscillationsGilbert2020,ominatoValleyDependentSpinTransport2020,yamamotoSpinCurrentMagnetic2021,ominatoAnisotropicSuperconductingSpin2021,yama2021Spin,Yama2022,ominato2022Ferromagnetica} 
	\begin{align}
	    G_{\bm 0}^R(\omega) = \frac{2S/\hbar}{(\omega-\omega_{\bm 0})+i(\alpha + \delta \alpha ) \omega}.
	    \label{magnon}
	\end{align}
	Here, we introduced a term, $i\alpha \omega$, in the denominator to express the spin relaxation within a bulk FI, where $\alpha$ indicates the strength of the Gilbert damping.
    The enhancement of the damping, $\delta \alpha$, is due to the adjacent Dirac electron system, calculated by
	\begin{align}
	    \delta \alpha = \frac{2S}{\hbar \omega} \sum_{\bm q} |\mathcal T_{\bm q, \bm 0}|^2 \text{Im}\, \chi_{\bm q}^R(\omega),
	    \label{eq:deltaalphadef}
	\end{align}
	where $\chi^R_{\bm q}(\omega)$ is the retarded component of the spin susceptibility (defined below). 
	We assume that the FMR peak described by ${\rm Im} \, G_{{\bm k}={\bm 0}}^R(\omega)$ is sufficiently sharp, i.e., $\alpha + \delta \alpha \ll 1$. 
	Then, the enhancement of the Gilbert damping can be regarded as almost constant around the peak ($\omega \simeq \omega_0$), allowing us to replace $\omega$ in $\delta \alpha$ with $\omega_{\bm 0}$.
	
\begin{figure}
        \centering
        \includegraphics[width=40mm]{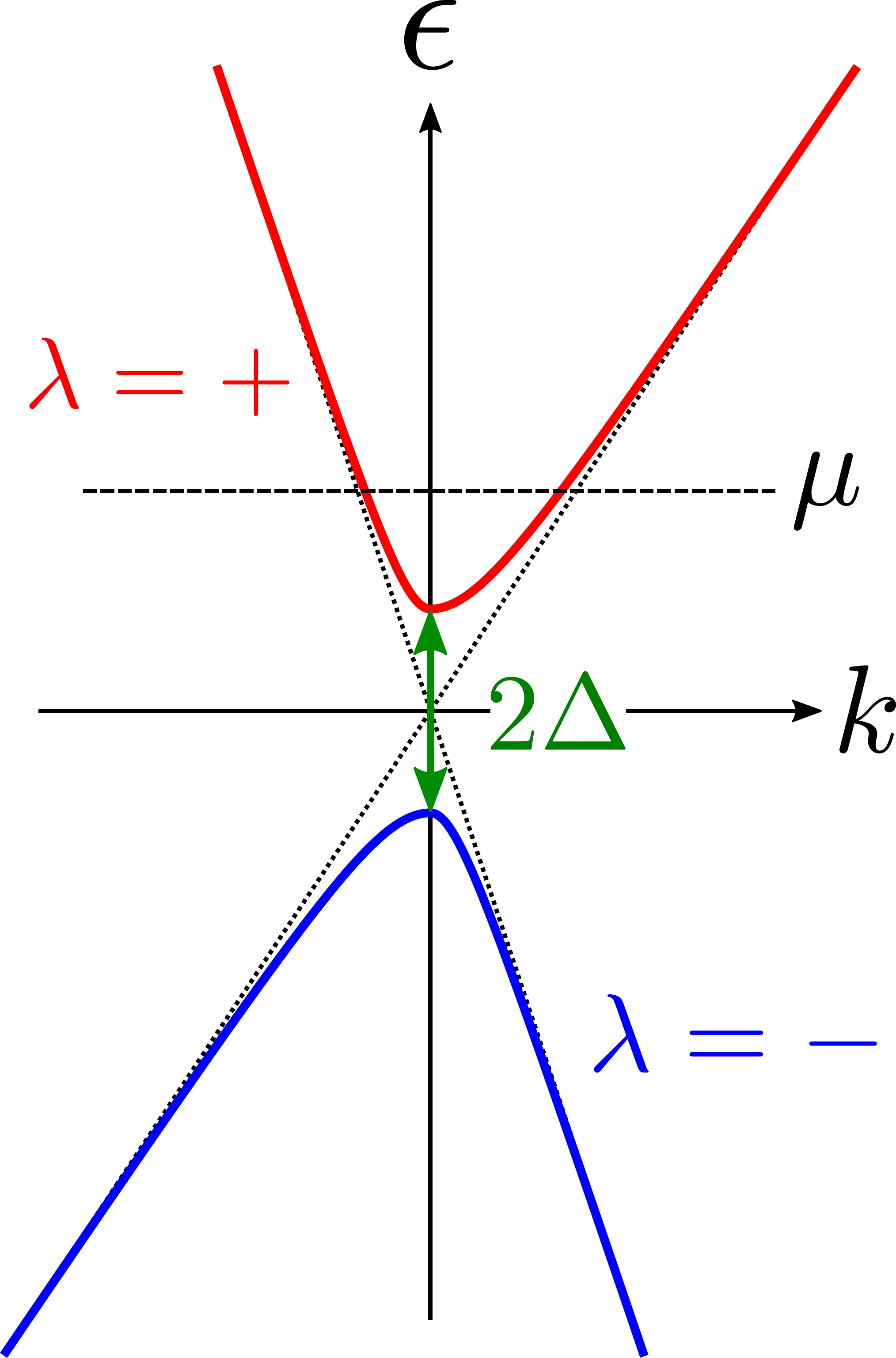}
        \caption{Schematic illustration of the band structure of the anisotropic Dirac electron system.
        The red band represents the conduction band with $\lambda =+$, and the blue band represents the valence band with $\lambda =-$.
        The chemical potential is in the conduction band.}
        \label{band}
    \end{figure}

	The retarded component of the spin susceptibility for the Dirac electrons:
	\begin{align}
	    \chi^R_{\bm q}(\omega) = \frac{i}{\hbar} \int^{\infty}_{-\infty} dt e^{i(\omega+i\delta)t} \theta(t)\average{[s^+_{\bm q}(t),s_{-\bm q}^-]} .
	\end{align}
	The retarded component of the spin susceptibility is derived from the following Matsubara Green function through analytic continuation $i\omega_l\to \hbar \omega + i\delta$: 
	\begin{align}
	    \chi_{\bm q}(i\omega_l) = \int^{\beta}_0 d\tau e^{i\omega_l\tau} \average{\hat s^+_{\bm q}(\tau)\hat s^-_{-\bm q}},
	\end{align}
	where $\omega_l=2\pi l/\beta$ is the bosonic Matsubara frequency with $l$ being integers.
	According to Wick's theorem, the Matsubara representation of the spin susceptibility is given by
	\begin{align}
	\label{eq:chiqomega}
	&\chi_{\bm q}(i\omega_l) \nonumber \\
	&= - \beta ^{-1} \sum_{\bm k,i\epsilon_n} \text{tr}[s^+g_{\bm k+\bm q}(i\epsilon_n+i\omega_l)s^-g_{\bm k}(i\epsilon_n)],
	\end{align}
	where $\sum_{i\epsilon_n}$ indicates the sum with respect to the fermionic Matsubara frequency, $\epsilon_n=(2\pi+1)n/\beta$.
	The imaginary part of the spin susceptibility is given by
	\begin{align}
	\label{eq:Imchi}
	    &\text{Im}\, \chi^R_{\bm q}(\omega) 
	    = -\pi \mathcal F(\theta,\phi)  \sum_{\bm k}\sum_{\lambda,\lambda'=\pm}  \left[  \frac{1}{2}+\frac{\lambda \lambda'}{6}\frac{2\Delta^2+\epsilon_{\bm k}^2}{\epsilon_{\bm k}\epsilon_{\bm k+\bm q}} \right]
	    \nonumber \\
	    & \hspace{4mm} \times
	    \Bigl[ f(\lambda' \epsilon_{\bm k+\bm q}) - f(\lambda \epsilon_{\bm k})\Bigr] 
	    \delta(\hbar \omega  -\lambda' \epsilon_{\bm k+\bm q}+\lambda \epsilon_{\bm k}),
	\end{align}
	where $f(\epsilon)=(e^{\beta(\epsilon-\mu)}+1)^{-1}$ is the Fermi distribution function, $\lambda=\pm$ is a band index (see Fig.~\ref{band}), and 
	$\mathcal F(\theta, \phi)$ is the dimensionless function which depends on the direction of the ordered localized spin, defined by
	\begin{align}
	    \mathcal F(\theta,\phi) = \left( \frac{2m}{\Delta} \right)^2 \sum_{\alpha}a_i\mathcal M_{i\alpha}a^*_j\mathcal M_{j\alpha}.
	\end{align}
	For detailed derivation, see Appendix~\ref{app:detailedderivation}.
	
	In this paper, we model the interfacial spin transfer as a combination of the clean and dirty processes. The former corresponds to the momentum-conserved spin transfer and the latter to the momentum-nonconserved one~\cite{ominatoAnisotropicSuperconductingSpin2021,ominato2022Ferromagnetica}.
	By averaging over the position of the localized spin at the interface, we can derive the matrix elements of the interfacial spin-transfer process as
	\begin{align}
	    |\mathcal T_{\bm q,\bm 0}|^2 = \mathcal T_1^2 \delta_{\bm q,\bm 0} + \mathcal T_2^2,
	\end{align}
	where $\mathcal T_1$ and $\mathcal T_2$ are the averaged matrix elements contributing to the clean and dirty processes, respectively. 
	Then, the enhancement of the Gilbert damping is given by
	\begin{gather}
	\delta \alpha = \frac{2S}{\hbar \omega}  \mathcal F(\theta, \phi)
	\Bigl\{
	    \mathcal T_1 \, \text{Im} \, \tilde \chi^R_{\text{uni}}(\omega_{{\bm 0}})
	    +
	    \mathcal T_2 \, \text{Im}\, \tilde \chi^R_{\text{loc}}(\omega_{{\bm 0}}) 
	\Bigr\},
	\end{gather}
	where $\chi^R_{\text{uni}}(\omega)$ and $\chi^R_{\text{uni}}(\omega)$ are the local and uniform spin susceptibilities defined by
	\begin{align}
	    \tilde \chi^R_{\text{loc}}(\omega_{{\bm 0}}) &= \mathcal F^{-1}(\theta,\phi) \sum_{\bm q} \chi^R_{\bm q}(\omega_{{\bm 0}}),	    \\
	    \tilde \chi^R_{\text{uni}}(\omega_{{\bm 0}}) &= \mathcal F^{-1}(\theta,\phi) \chi^R_{\bm 0}(\omega_{{\bm 0}}),
	\end{align}
	respectively. From Eq.~(\ref{eq:Imchi}), their imaginary parts 
	are calculated as
	\begin{align}
	    \text{Im}\, \tilde \chi_{\text{loc}}^R(\omega_{{\bm 0}}) 
	    &= - \pi n_D^2 \int d\epsilon \nu(\epsilon) \nu(\epsilon+\hbar \omega_{{\bm 0}}) 
	    \nonumber \\ 
	    & \hspace{2mm} \times \left[  \frac{1}{2}+\frac{2\Delta^2+\epsilon^2}{6\epsilon(\epsilon+\hbar \omega_{{\bm 0}})} \right] \Bigl[ f(\epsilon +\hbar \omega_{{\bm 0}}) - f(\epsilon)\Bigr], \label{eq:Imchiloc}
	    \\
	    \text{Im}\, \tilde \chi^{\text R}_{\text{uni}}(\omega_{{\bm 0}})
	    &=  -\pi n_D  \nu \left(\tfrac{\hbar\omega_{{\bm 0}}}{2}\right) \frac{\hbar^2 \omega_{{\bm 0}}^2-4\Delta^2}{3\hbar^2 \omega_{{\bm 0}}^2} \nonumber \\
	    & \hspace{2mm} \times \Bigl[ f(\tfrac{\hbar\omega_{{\bm 0}}}{2}) - f(-\tfrac{\hbar\omega_{{\bm 0}}}{2})\Bigr].
	     \label{eq:Imchiuni}
	\end{align}
	The enhancement of the Gilbert damping, $\delta \alpha$, depends on the direction of the ordered localized spin through the dimensionless function $\mathcal F(\theta,\phi)$ regardless of the interfacial condition.
    
	\begin{figure}
        \centering
        \includegraphics[width=65mm]{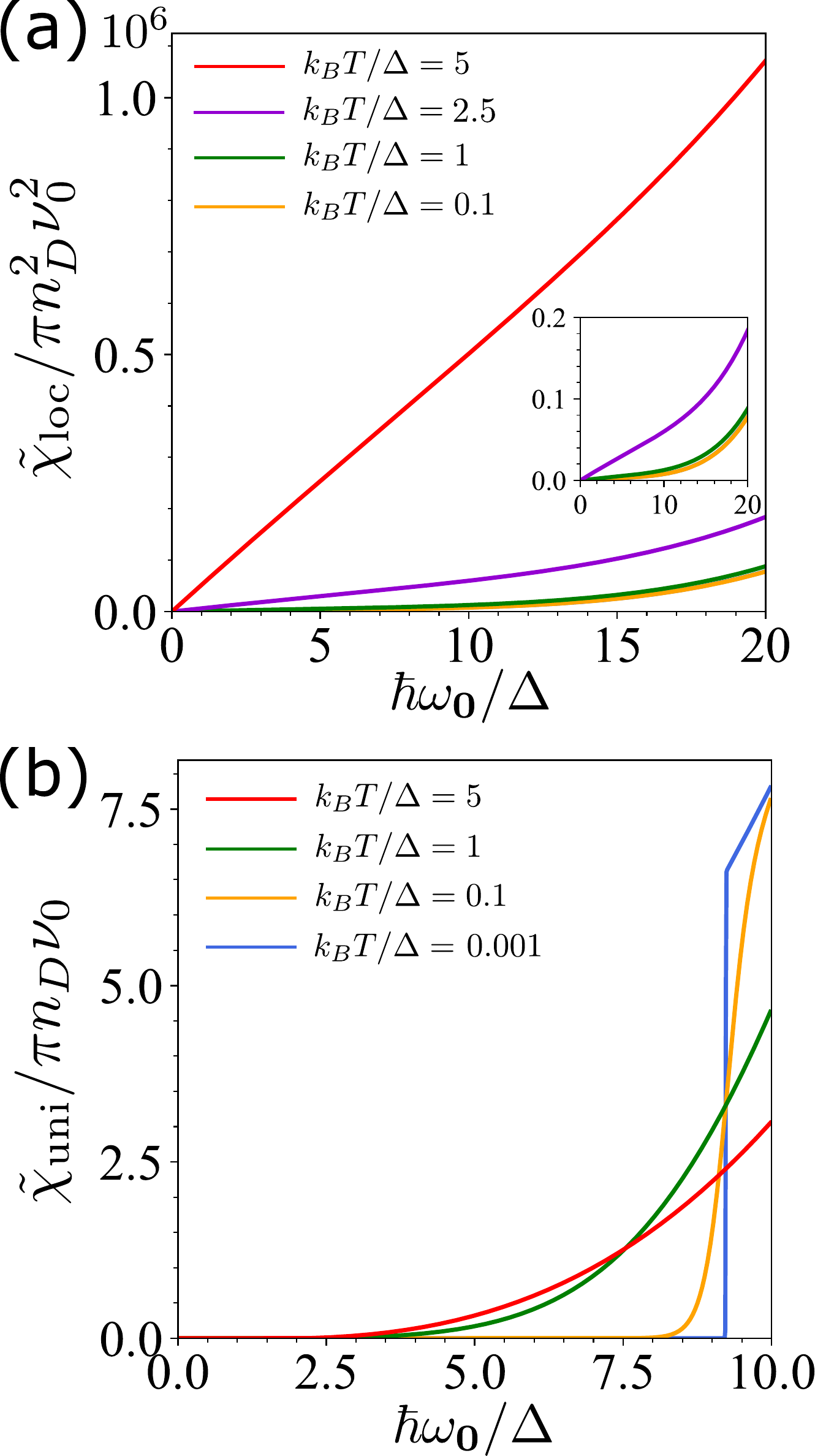}
        \caption{
        FMR frequency dependence of the (a) local and (b) uniform spin susceptibilities.
        The local spin susceptibility is normalized by $\pi n_D^2 \nu_0^2$ and scaled by $10^6$, and the uniform spin susceptibility is normalized by $\pi n_D \nu_0$ with $\nu_0\equiv 1/2\pi ^2\hbar^3 \sqrt{\text{det}\alpha_{ij}}$.
        Note that $k_B$ is the Boltzmann constant.
        The line with $k_BT/\Delta =0.001$ is absent in (a) because the local spin susceptibility approaches zero at low temperature. 
        }
        \label{fig.freq_dep}
    \end{figure}

    By contrast, the FMR frequency dependence of $\delta \alpha$ reflects the interfacial condition;
    for a clean interface, it is determined mainly by $\text{Im}\, \chi^R_{\text{uni}}(\omega_{{\bm 0}})$, whereas for a dirty interface, it is determined by $\text{Im}\, \chi^R_{\text{loc}}(\omega_{{\bm 0}})$.
    The FMR frequency dependence of the local and uniform spin susceptibilities, $\text{Im}\, \chi^R_{\text{loc}}(\omega_{{\bm 0}})$ and $\text{Im}\, \chi^R_{\text{uni}}(\omega_{{\bm 0}})$, are plotted in Figs.~\ref{fig.freq_dep}~(a) and (b), respectively.
    The local and uniform spin susceptibilities are normalized by $\pi n_D^2 \nu_0^2$ and $\pi n_D \nu_0$, respectively, where $\nu_0\equiv 1/2\pi ^2\hbar^3 \sqrt{\text{det}\alpha_{ij}}$ is defined.
    In the calculation, the ratio of the chemical potential to the energy gap was set to $\mu/\Delta \simeq 4.61$, which is the value in the bismuth~\cite{fuseyaTransportPropertiesDiamagnetism2015}.
    According to Fig.~\ref{fig.freq_dep}~(a), the local spin susceptibility increases linearly with the frequency $\omega$ in the low-frequency region. 
    This $\omega$-linear behavior can be reproduced analytically for low temperatures and $\hbar \omega \ll \mu$:
	\begin{align}
	    \text{Im}\, \tilde \chi_{\text{loc}} (\omega_{\bm 0})
	    &\simeq \hbar \omega_{{\bm 0}} \frac{\pi}{2} n_D^2  [\nu(\mu)]^2 \left[ 1+\frac{2\Delta^2+\mu^2}{3\mu^2} \right].
	\end{align}
	Fig.~\ref{fig.freq_dep}~(b) indicates a strong suppression of the uniform spin susceptibility below a spin-excitation gap ($\omega_{\bm 0} < 2\mu$).
	This feature can be checked by its analytic form at zero temperature:
	\begin{align}
	    \text{Im}\, \tilde \chi^{\text R}_{\text{uni}}(\omega_{{\bm 0}})
	    &= \pi n_D  \nu \left(\tfrac{\hbar\omega_{{\bm 0}} }{2}\right) \frac{\hbar^2 \omega_{{\bm 0}} ^2-4\Delta^2}{3\hbar^2 \omega_{{\bm 0}} ^2}\theta(\hbar \omega_{{\bm 0}} -2\mu) .
	\end{align}
	Thus, the FMR frequency dependence of the enhancement of the Gilbert damping depends on the interfacial condition.
	This indicates that the measurement of the FMR frequency dependence may provide helpful information on the randomness of the junction.

	\section{Result\label{sec.result}}

\begin{figure}
        \centering
        \includegraphics[width=75mm]{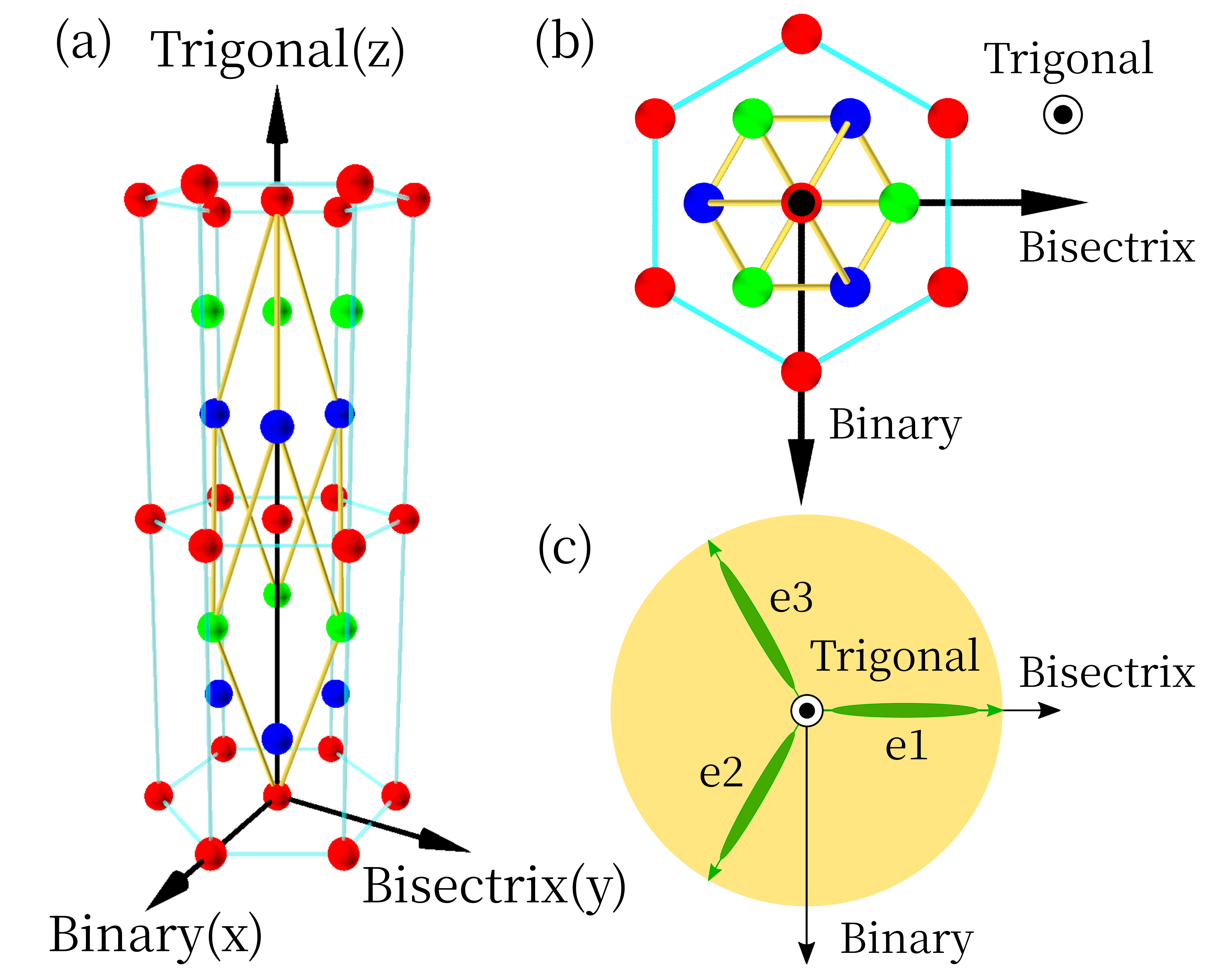}
        \caption{(a) The rhombohedral lattice structure of bismuth.
        The $x$-axis, $y$-axis, and $z$-axis are chosen as the binary axis with $C_2$ symmetry, the bisectrix axis, and the trigonal axis with $C_3$ symmetry, respectively.
        The yellow lines represents the unit cell of the rhombohedral lattice.
        (b) The rhombohedral structure viewed from the trigonal axis.
        (c) Schematic illustration of the band structure at the Fermi surface.
        The three electron ellispoids at $L$-points are dominant contribution to the spin transport. 
        }
        \label{rhombohedral}
    \end{figure}

	We consider bismuth, which is one of the anisotropic Dirac electron systems~\cite{edelmanElectronsBismuth1976,tangElectronicPropertiesNanostructured2014,fuseyaSpinHallEffectDiamagnetism2014,fuseyaTransportPropertiesDiamagnetism2015,zhu2011Angleresolved}. 
	The crystalline structure of pure bismuth is a rhombohedral lattice with the space group of $R\bar 3m$ symmetry, see Figs.~\ref{rhombohedral}~(a) and (b).
	It is reasonable to determine the Cartesian coordinate system in the rhombohedral structure using the trigonal axis with $C_3$ symmetry, the binary axis with $C_2$ symmetry, and the bisectrix axis, which is perpendicular to the trigonal and binary axes.
	Hereafter, we choose the $x$-axis as the binary axis, the $y$-axis as the bisectrix axis, and the $z$-axis as the trigonal axis.
	Note that the trigonal, binary, and bisectrix axes are denoted as $[0001]$, $[1\bar 210]$, and $[10\bar 10]$, respectively, where the Miller-Bravais indices are used.
	The bismuth's band structure around the Fermi surface consists of three electron ellipsoids at $L$-points and one hole ellipsoid at the $T$-point. 
	It is well known that the electron ellipsoids are the dominant contribution to the transport phenomena since electron's mass is much smaller than that of the hole, see Fig.~\ref{rhombohedral}~(c).
	Therefore, the present study considers only the electron systems at the $L$-points.
	The electron ellipsoids are  significantly elongated, with the ratio of the major to minor axes being approximately $15:1$. 
	Each of the three electron ellipsoids can be converted to one another with $2\pi/3$ rotation around the trigonal axis. 
	The electron ellipsoid along the bisectrix axis is labeled as $e1$, and the other two-electron ellipsoids are labeled $e2$ and $e3$. 
	The inverse mass tensor for the $e1$ electron ellipsoids is given by
	\begin{align}
	    \tensor{\alpha}_{e1} = \begin{pmatrix}
    \alpha_1 & 0 & 0 \\[5pt]
    0 & \alpha_2 & \alpha_4 \\[5pt]
    0 & \alpha_4 & \alpha_3
    \end{pmatrix}.
	\end{align}
	The inverse mass tensor of the electron ellipsoids $e2$ and $e3$ are obtained by rotating that of $e1$ by $2\pi/3$ rotation as below: 
	\begin{align}
	\tensor{\alpha}_{e2,e3} = \frac{1}{4} \begin{pmatrix}
    \alpha_1 + 3\alpha_2 & \pm \sqrt 3(\alpha_1-\alpha_2) & \pm 2\sqrt 3\alpha_4 \\[5pt]
    \pm \sqrt 3(\alpha_1-\alpha_2) & 3\alpha_1 + \alpha_2 & -2\alpha_4 \\[5pt]
    \pm 2\sqrt{3}\alpha_4 & -2\alpha_4 & 4\alpha_3
    \end{pmatrix}.
    \end{align}
    
   Let us express the dimensionless function $\mathcal F(\theta,\phi)$ representing the localized spin direction dependence of the damping enhancement on the inverse mass tensors.
    \begin{gather}
    \mathcal F(\theta,\phi) = \left( \frac{2m}{\Delta} \right)^2 \sum_{\alpha} \Bigl[ (\sin^2\phi + \sin^2\theta \cos^2\phi)  \mathcal M_{x\alpha}^2 \nonumber  \\
    + (\cos^2\phi+\sin^2 \theta \sin^2 \phi) \mathcal M_{y\alpha}^2
    \nonumber \\
    + \cos^2\theta (\mathcal M_{z\alpha}^2 - \sin 2\phi \mathcal M_{x\alpha} \mathcal M_{y\alpha})
    \nonumber \\
    + \sin 2\theta \mathcal M_{z\alpha} (\mathcal M_{x\alpha} \cos \phi + \mathcal M_{y\alpha} \sin \phi)  \Bigr].
\end{gather}
Here, we use the following calculations:
    \begin{gather}
	    \sum_{\alpha} \mathcal M_{x\alpha}^2 = \frac{\Delta^2}{4} ( \alpha_{yy}\alpha_{zz}-\alpha_{yz}^2)_{\text{total}}
	    = \frac{\Delta^2}{4m^2}\bar \kappa_{\perp}
	    ,
	    \\
	    \sum_{\alpha} M_{y\alpha}^2 = \frac{\Delta^2}{4} (\alpha_{zz}\alpha_{xx}-\alpha_{zx}^2)_{\text{total}}
	    =\frac{\Delta^2}{4m^2} \bar \kappa_{\perp},
	    \\
	    \sum_{\alpha} M_{z\alpha}^2 = \frac{\Delta^2}{4} ( \alpha_{xx}\alpha_{yy}-\alpha_{xy}^2)_{\text{total}}
	    =\frac{\Delta^2}{4m^2} \bar \kappa_{\parallel},
	    \\
	    \sum_{\alpha} \mathcal M_{i\alpha} \mathcal M_{j\alpha} = \frac{\Delta^2}{4} (\alpha_{ik}\alpha_{jk}-\alpha_{ij}\alpha_{kk})_{\text{total}}=0,
	\end{gather}
	where $i,j,k$ are cyclic.
	$(\cdots)_{\text{total}}$ represents the summation of the contributions of the three electron ellipsoids, and $\bar \kappa_{\parallel}$, $\bar \kappa_{\perp}$ $(>0)$ are the total Gaussian curvature of the three electron ellipsoids normalized by the electron mass $m$, given by
    \begin{gather}
        \bar \kappa_{\parallel} = 3m^2 \alpha_1\alpha_2,\\
        \bar \kappa_{\perp} = \frac{3}{2}m^2[(\alpha_1+\alpha_2)\alpha_3-\alpha_4^2].
    \end{gather}
    Hence, the dimensionless function $\mathcal F$ is given by
	\begin{align}
        \mathcal F(\theta) =  (1+\sin^2\theta) \bar \kappa_{\perp} + \cos^2\theta \bar \kappa_{\parallel} .
    \end{align}
    
\begin{figure}
        \centering
        \includegraphics[width=65mm]{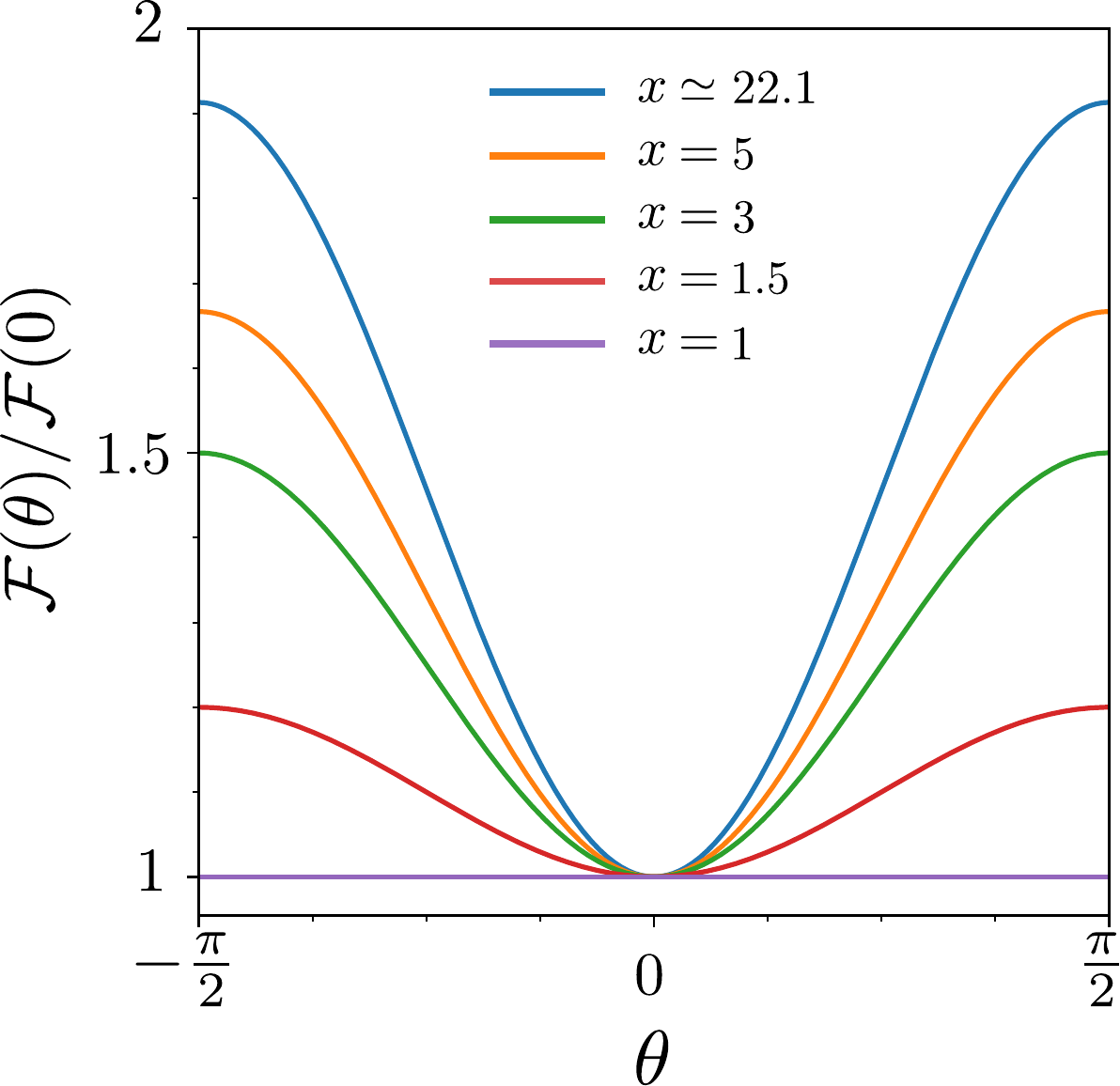}
        \caption{
        The $\theta$-dependence of the damping enhancement for different $x$.
        The ratio of the total Gaussian curvatures $x=\bar{\kappa}_{\perp}/\bar{\kappa}_{\parallel}$ represents the anisotropy of the band structure.
        The blue line with $x=22.1$ corresponds to the damping enhancement in single-crystalline bismuth, and the other lines correspond to that in the weakly anisotropic band structure.
        As can be seen from the graph, the $\theta$-dependence of the damping enhancement decreases as the more weakly anisotropic band structure, and the angular dependence turns out to vanish in an isotropic Dirac electron system with $x=1$.
        }
        \label{graph}
    \end{figure}

    The results suggest that the variation of the damping enhancement depends only on the polar angle $\theta$, which is the angle between the direction of the ordered localized spin $\average{\bm S}_0$ and the trigonal axis. 
    It is also found that the $\theta$ dependence of the damping enhancement originates from the anisotropy of the band structure.
    The dimensionless function $\mathcal F(\theta)$ is plotted in Fig.~\ref{graph} by varying the ratio of the total Gaussian curvatures $x=\bar{\kappa}_{\perp}/\bar{\kappa}_{\parallel}$, which corresponds to the anisotropy of the band structure.
    Figure~\ref{graph} shows that the $\theta$-dependence of the damping enhancement decreases with smaller $x$ and the angular dependence vanishes in an isotropic Dirac electron system $x=1$.
    Bismuth is known to have a strongly anisotropic band structure.
    The magnitude of the matrix elements of the inverse mass $\alpha_1$-$\alpha_4$ was experimentally determined as $m \alpha_1=806$, $m\alpha_2=7.95$, $m\alpha_3=349$, and $m \alpha_4=37.6$. The total Gaussian curvatures are evaluated as~\cite{fuseyaTransportPropertiesDiamagnetism2015}
    \begin{gather}
        \bar \kappa_{\parallel} \simeq 1.92\times 10^4,
        \\
        \bar \kappa_{\perp} \simeq 4.24\times 10^5.
    \end{gather}
    The ratio of the total Gaussian curvature is estimated as $x\simeq 22.1$.
    Therefore, the damping enhancement is expected to depend strongly on the polar angle $\theta$ in a bilayer system composed of single-crystalline bismuth and ferromagnetic insulator.
    Conversely, the $\theta$-dependence of the damping enhancement is considered to be suppressed for polycrystalline bismuth.
    
    The damping enhancement is independent of the azimuthal angle $\phi$. Therefore, it is invariant even on rotating the spin orientation around the trigonal axis.
    The reason is that the azimuthal angular dependence of the damping enhancement cancels out when the contributions of the three electron ellipsoids are summed over, although each contribution depends on the azimuthal angle.
    The azimuthal angular dependence of the damping enhancement is expected to remain when strain breaks the in-plane symmetry.
    Additionally, suppose the spin can be injected into each electron ellipsoid separately, 
    e.g., by interfacial manipulation of the bismuth atoms.
    In that case, the damping enhancement depends on the azimuthal angle of the spin orientation of the ferromagnetic insulator~\cite{ominatoValleyDependentSpinTransport2020}.
    This may be one of the probes of the electron ellipsoidal selective transport phenomena.
        
    It is also noteworthy that the damping enhancement varies according to the ordered localized spin direction with both clean and dirty interfaces; that is independent of 
    whether momentum is conserved in interfacial spin transport. 
    Conversely, it was reported that the spin orientation dependence of the damping enhancement due to the Rashba and Dresselhaus spin-orbit interaction turned out to vanish by interfacial inhomogeneity~\cite{yama2021Spin,Yama2022}.

    \section{conclusion\label{sec.conclusion}}
    
    We theoretically studied spin pumping from a ferromagnetic insulator to an anisotropic Dirac electron system.
    We calculated the enhancement of the Gilbert damping in the second perturbation concerning the proximity interfacial exchange interaction by considering the interfacial randomness. For illustration, we calculated the enhancement of the Gilbert damping for an anisotropic Dirac system realized in bismuth.
    We showed that the Gilbert damping varies according to the polar angle between the ordered spin $\average{\bm S}_0$ and the trigonal axis of the Dirac electron system whereas it is invariant in its rotation around the trigonal axis.
    Our results indicate that the spin pumping experiment can provide helpful information on the anisotropic band structure of the Dirac electron system.

    The Gilbert damping is invariant in the rotation around the trigonal axis because the contributions of each electron ellipsoid depend on the in-plane direction of the ordered spin $\average{\bm S}_0$.
    Nevertheless, the total contribution becomes independent of the rotation of the trigonal axis after summing up the contributions from the three electron ellipsoids that are related to each other by the $C_3$ symmetry of the bismuth crystalline structure.
    If the spin could be injected into each electron ellipsoid separately, it is expected that the  in-plane direction of the ordered localized spin would influence the damping enhancement. 
    This may be one of the electron ellipsoid selective spin injection probes.
    The in-plane direction's dependence will also appear when a static strain is applied.
    A detailed discussion of these effects is left as a future problem.

	\begin{acknowledgments}
    The authors would like to thank A. Yamakage and Y. Ominato for helpful and enlightening discussions.
    The continued support of Y. Nozaki is greatly appreciated.
    We also thank H, Nakayama for the daily discussions.
	This work was partially supported by JST CREST Grant No. JPMJCR19J4, Japan.
	This work was supported by JSPS KAKENHI for Grants (Nos. 20H01863, 20K03831, 21H04565, 21H01800, and 21K20356).
	MM was supported by the Priority Program of the Chinese Academy of Sciences, Grant No. XDB28000000.
	\end{acknowledgments}
	\appendix
	
	\section{Magnetic moment of electrons in Dirac electron system\label{spin_op}}
	
	In this section, we define the spin operators in the Dirac electron systems.
	The Wolff Hamiltonian around the L point is given by $\mathcal H_{\text D}=\rho_3\Delta -\rho_2  \bm \pi \cdot \bm v$, where $v_i = \sum_{\alpha}w_{i\alpha}\sigma ^{\alpha}$ with $w_{i\alpha}$ being the matrix component of the velocity vectors and $\bm \pi= \bm p+\frac{e}{c}\bm A$ is the momentum operator including the vector potential.
	It is reasonable to determine the magnetic moment of electrons in an effective Dirac system as the coefficient of the Zeeman term.
	The Wolff Hamiltonian is diagonalized by the Schrieffer-Wolff transformation up to $\bm v/\Delta$ as below:
	\begin{align}
	    &e^{i\xi}\mathcal H_{\text D} e^{-i\xi} \simeq \left[ \Delta + \frac{1}{2\Delta}(\bm \pi \cdot \bm v)^2  \right] \rho_3, 
	\end{align}
	where $\xi=\frac{\rho_1}{2\Delta} \bm \pi \cdot \bm v$ is chosen to erase the off-diagonal matrix for the particle-hole space.
	We can proceed calculation as follows:
	\begin{align}
	    (\bm \pi \cdot \bm v)^2 
	    &= \pi_i \pi_{j} w_{i\alpha}w_{j\beta} (\delta_{\alpha \beta} + i\epsilon_{\alpha \beta \gamma}\sigma ^{\gamma}), \nonumber \\
	    &= (\pi_i w_{i\alpha})^2 + \frac{i}{2}\epsilon_{\alpha \beta \gamma} \sigma^{\gamma} [\bm \pi \times \bm \pi]_i \epsilon_{ijk} w_{j\alpha}w_{k\beta},
	    \nonumber \\
	    &= \Delta \left( \bm \pi \cdot \alpha \cdot \bm \pi + \frac{\hbar e }{c\Delta} \mathcal M_{i\alpha} \sigma^{\alpha} B_i 
	    \right),
	\end{align}
	where we used $(\bm \pi\times \bm \pi)=\frac{e\hbar}{ci}\nabla \times \bm A$ and $M_{i\alpha}$ is defined as
	\begin{gather}
	    \mathcal M_{i\alpha} =\frac{1}{2} \epsilon _{\alpha \beta \gamma} \epsilon_{ijk}w_{j\beta} w_{k\gamma}.
	\end{gather}
	Finally, we obtain
	\begin{align}
	    &e^{i\xi}\mathcal H_{\text D} e^{-i\xi}  
	    \simeq \left[ \Delta + \frac{\bm \pi \cdot \tensor{\alpha} \cdot \bm \pi}{2} \right] - B_i \mu_{{\rm s},i},
	\end{align}
    where $\mu_{{\rm s},i}$ is a magnetic moment of the Dirac electrons defined as
	\begin{gather}
	    \mu_{\text s,i} = -\frac{\hbar e}{2c\Delta}  \mathcal M_{i\alpha} \rho_3 \sigma^{\alpha} = - \frac{\hbar e}{2c\Delta }  \mathcal M_{i\alpha}
	    \begin{pmatrix}
	    \sigma^{\alpha} & 0\\
	    0 & -\sigma^{\alpha} 
	    \end{pmatrix}.
	\end{gather}
	In the main text, we defined the spin operator $\bm s$ as the magnetic moment $\bm \mu_s$ divided by the Bohr magnetization $\mu_B=\hbar e/2mc$, i.e.,
	\begin{align}
	    s^i = -\frac{\mu_{{\rm s},i}}{\mu_{\rm B}}
	    = \frac{m}{\Delta }  \mathcal M_{i\alpha}
	    \begin{pmatrix}
	    \sigma^{\alpha} & 0\\
	    0 & -\sigma^{\alpha} 
	    \end{pmatrix}.
	    \label{app:spinop}
	\end{align}
	For an isotropic Dirac system, the matrix component is given by $w_{i\alpha}=v\delta_{i\alpha}$ and Eq.~(\ref{app:spinop}) reproduces the well-known form of the spin operator
	\begin{align}
	    \bm s = \frac{g^*}{2} 
	    \begin{pmatrix}
	    \bm \sigma & 0\\
	    0 & -\bm \sigma
	    \end{pmatrix},
	\end{align}
	where $g^*=2m/m^*$ is the effective $g$-factor with $m^*=\Delta/v^2$ being effective mass.
	
	\section{Linear Response Theory}
	\label{app:linearresponse}
	In this section, we briefly explain how the microwave absorption rate is written in terms of the uniform spin correlation function.
	The Hamiltonian of an external circular-polarized microwave is written as
    \begin{align}
	    \hat{\mathcal H}_{\text{rf}} &= -\frac{g\mu_B h_{\text{rf}}}{2} \sum_i (S_i^- e^{-i\omega t} + S_i^+ e^{i\omega t}) \nonumber \\
	    &= - \frac{g\mu_B h_{\text{rf}}\sqrt{n_{\rm F}}}{2} (S_{\bm 0}^{-} e^{-i\omega t} + S_{\bm 0}^{+} e^{i\omega t}),
    \end{align}	
	where $h_{\text{rf}}$ is an amplitude of the magnetic field of the microwave, $S_{\bm k}^{\pm}$ are the Fourier transformations defined as
    \begin{align}
        S_{\bm k}^{\pm} = \frac{1}{\sqrt{n_{\rm F}}} \sum_i S_i^\pm e^{-i {\bm k} \cdot {\bm R}_i},
    \end{align}
	and ${\bm R}_i$ is the position of the locazed spin $i$.
    Using the linear response theory with respect to $\hat{\mathcal H}_{\text{rf}}$, the expectation value of the local spin is calculated as
	\begin{align}
	    \average{S^+_{\bm 0}}_{\omega}=G^R_{\bm 0}(\omega) \times \frac{g\mu_B h_{\text{rf}}\sqrt{n_{\rm F}}}{2},
	\end{align}
    where $G^R_{\bm k}(\omega)$ is the spin correlation function defined in Eq.~(\ref{eq:GRomegadef}).
    Since the microwave absorption is determined by the dissipative part of the response function, it is proportional to ${\rm Im} \, G^R_{\bm 0}(\omega)$, that reproduces a Lorentzian-type FMR lineshape.
    As explained in the main text, the change of the linewidth of the microwave absorption, $\delta \alpha$, gives information on spin excitation in the Dirac system via the spin susceptibility as shown in Eq.~(\ref{eq:deltaalphadef}).

\begin{widetext}
	
	\section{Spin susceptibility of Dirac electrons}
	\label{app:detailedderivation}
	
    In this section, we give detailed derivation of Eq.~(\ref{eq:Imchi}).
	The trace part in Eq.~(\ref{eq:chiqomega}) is calculated as
	\begin{align}
	    \text{tr}[s^+g_{\bm k+\bm q}(i\epsilon_n+i\omega_l)s^-g_{\bm k}(i\epsilon_n)]
	    = \frac{
	    [
	        (i\epsilon_n+i\omega_l +\mu)(i\epsilon _n +\mu) + \Delta^2
	    ] \text{tr}[s^+s^-] -\text{tr}[s^+\hbar (\tilde{\bm k}+\tilde{\bm q})\cdot \bm \sigma s^-\hbar \tilde{\bm k}\cdot \bm \sigma]
	    }{
	    [(i\epsilon_n+i\omega_l+\mu)^2-\epsilon_{\bm k+\bm q}^2][(i\epsilon_n+\mu)^2-\epsilon_{\bm k}^2]
	    }
	    ,
	    \label{trace}
	\end{align}
	where $(\tilde{\bm k}+\tilde{\bm q})\cdot \bm \sigma = (\bm k+\bm q)\cdot \bm v$.
	Using the following relations
	\begin{align}
	    \text{tr}[s^+s^-] &= \left( \frac{2m}{\Delta} \right)^2 \sum_{\alpha} a_i\mathcal M_{i\alpha} a_j^*\mathcal M_{j\alpha},
	    \\
	    \text{tr}[s^+\hbar (\tilde{\bm k}+\tilde{\bm q}) \cdot \bm \sigma s^-\hbar \tilde{\bm k}\cdot \bm \sigma] &= \left( \frac{2m}{\Delta} \right)^2
	    \sum_{\alpha} ( 2a_i\mathcal M_{i\alpha} \hbar \tilde k_{\alpha} a_j^*\mathcal M_{j\beta} \hbar \tilde k_{\beta}
	    - \hbar ^2\tilde k^2 a_i\mathcal M_{i\alpha} a_j^*\mathcal M_{j\alpha} 
	    ),
	\end{align}
	the spin susceptibility is given by
	\begin{align}
	    \chi_{\bm q}(i\omega_l) = - 2\mathcal F(\theta,\phi) \sum_{\bm k} \beta^{-1} \sum_{i\epsilon_n} \frac{ (i\epsilon_n+i\omega_l+\mu) (i\epsilon_n+\mu) + \Delta^2 + \hbar^2 \tilde k^2/3 }{[(i\epsilon_n+i\omega_l+\mu)^2-\epsilon_{\bm k+\bm q}^2][(i\epsilon_n+\mu)^2-\epsilon_{\bm k}^2]}\blue{,}
	\end{align}
	where we dropped the terms proportional to $\tilde{k}_\alpha \tilde{k}_\beta$ ($\alpha \ne \beta$) because they vanish after the summation with respect to the wavenumber ${\bm k}$.
	Here, we introduced a dimensionless function, $\mathcal F(\theta,\phi)= \left( 2m/\Delta \right)^2 \sum_{\alpha} a_i\mathcal M_{i\alpha} a_j^*\mathcal M_{j\alpha}$, which depends on the direction of the magnetization of the FI.
	Representing the Matsubara summation as the following contour integral, we derive
	\begin{align}
	    \chi_{\bm q}(i\omega_l) &= -2\mathcal F(\theta,\phi)  \sum_{\bm k} \oint \frac{dz}{4\pi i} \tanh\left( \frac{\beta (z-\mu)}{2} \right) \frac{ z(z+i\omega_l) + \Delta^2 + \hbar^2 \tilde k^2/3 }{[(z+i\omega _l)^2-\epsilon_{\bm k+\bm q}^2][z^2-\epsilon_{\bm k}^2]},\\
	    &= 2 \mathcal F(\theta,\phi)  \sum_{\bm k} \oint \frac{dz}{2\pi i} f(z) \frac{ z(z+i\omega_l) + \Delta^2+\hbar^2 \tilde k^2/3}{[(z+i\omega _l)^2-\epsilon_{\bm k+\bm q}^2][z^2-\epsilon_{\bm k}^2]},
	\end{align}
	We note that $\tanh(\beta(z-\mu)/2)$ has poles at $z=i\epsilon_n+\mu$ and is related to the Fermi distribution function $f(z)$ as $\tanh[\beta (z-\mu)/2]=1-2f(z)$.
	Using the following identities
	\begin{align}
	    \frac{1}{z^2-\epsilon_{\bm k}^2} &
	    = \frac{1}{2\epsilon_{\bm k}} \sum_{\lambda=\pm} \frac{\lambda}{z-\lambda \epsilon_{\bm k}},\\
	    \frac{z}{z^2-\epsilon_{\bm k}^2} 
	    &= \frac{1}{2} \sum_{\lambda=\pm} \frac{1}{z-\lambda \epsilon_{\bm k}},
	\end{align}
	the spin susceptibility is given by
	\begin{align}
	    \chi_{\bm q}(i\omega_l) &= \mathcal F(\theta,\phi)  \sum_{\bm k} \oint \frac{dz}{2\pi i} f(z) \sum_{\lambda ,\lambda'=\pm} \left[
	     \frac{1}{2} + \frac{(\Delta^2+
	     \hbar^2 \tilde k^2/3)\lambda \lambda'}{2\epsilon_{\bm k}\epsilon_{\bm k+\bm q}}
	    \right]
	    \frac{1}{z -\lambda \epsilon_{\bm k}} \frac{1}{z+i\omega_l -\lambda' \epsilon_{\bm k+\bm q}} 
	    ,
	    \\
	    &= \mathcal F(\theta,\phi)  \sum_{\bm k} \sum_{\lambda,\lambda'=\pm} \left[  \frac{1}{2}+\frac{\lambda \lambda'}{6}\frac{2\Delta^2+\epsilon_{\bm k}^2}{\epsilon_{\bm k}\epsilon_{\bm k+\bm q}} \right] \frac{f(\lambda'\epsilon_{\bm k+\bm q}) - f(\lambda \epsilon_{\bm k})}{i\omega_l -\lambda'\epsilon_{\bm k+\bm q}+\lambda \epsilon_{\bm k}}
	    .
	\end{align}
	By the analytic continuation $i\omega_l=\hbar \omega +i\delta$, we derive the retarded spin susceptibility as below:
	\begin{align}
	    \chi^R_{\bm q}(\omega) = \mathcal F(\theta,\phi)  \sum_{\bm k} \sum_{\lambda,\lambda'=\pm} \left[  \frac{1}{2}+\frac{\lambda \lambda'}{6}\frac{2\Delta^2+\epsilon_{\bm k}^2}{\epsilon_{\bm k}\epsilon_{\bm k+\bm q}} \right] \frac{f(\lambda'\epsilon_{\bm k+\bm q}) - f(\lambda \epsilon_{\bm k})}{\hbar \omega +i\delta -\lambda'\epsilon_{\bm k+\bm q}+\lambda \epsilon_{\bm k}}.
	\end{align}
	The imaginary part of the spin susceptibility is given by
	\begin{align}
	    \text{Im}\chi^R_{\bm q}(\omega) = -\pi \mathcal F(\theta,\phi)  \sum_{\bm k}\sum_{\lambda,\lambda'=\pm}  \left[  \frac{1}{2}+\frac{\lambda \lambda'}{6}\frac{2\Delta^2+\epsilon_{\bm k}^2}{\epsilon_{\bm k}\epsilon_{\bm k+\bm q}} \right]
	    \Bigl[ f(\lambda' \epsilon_{\bm k+\bm q}) - f(\lambda \epsilon_{\bm k})\Bigr] 
	    \delta(\hbar \omega  -\lambda' \epsilon_{\bm k+\bm q}+\lambda \epsilon_{\bm k}).
	\end{align}
    From this expression, Eqs.~(\ref{eq:Imchiloc}) and (\ref{eq:Imchiuni}) for the imaginary parts of the uniform and local spin susceptibilities can be obtained by replacing the sum with respect to ${\bm k}$ and $\lambda$ with an integral over the energy $\epsilon$ as follows:	
	\begin{align}
	    n_{\text D}^{-1} \sum_{\bm k,\lambda} \mathcal A(\lambda \epsilon_{\bm k}) &\rightarrow \int\frac{d^3\tilde k}{(2\pi)^3 \sqrt{\Delta^3 \text{det}\alpha_{ij}}} \mathcal A(\lambda \epsilon_{\bm k}) =
	    \int^{\infty}_{-\infty} d\epsilon \nu(\epsilon)\mathcal A(\epsilon)
	    ,
	\end{align}
	where $\mathcal A$ is an arbitrary function. 
	Note that the Jacobian of the transformation from $\bm k$ to $\tilde{\bm k}$ is given by $\text{det}(dk_i/d\tilde k_j)=1/\sqrt{\Delta^3 \text{det}\alpha_{ij}}$.   
	
\end{widetext}


\begin{thebibliography}{61}%
\makeatletter
\providecommand \@ifxundefined [1]{%
 \@ifx{#1\undefined}
}%
\providecommand \@ifnum [1]{%
 \ifnum #1\expandafter \@firstoftwo
 \else \expandafter \@secondoftwo
 \fi
}%
\providecommand \@ifx [1]{%
 \ifx #1\expandafter \@firstoftwo
 \else \expandafter \@secondoftwo
 \fi
}%
\providecommand \natexlab [1]{#1}%
\providecommand \enquote  [1]{``#1''}%
\providecommand \bibnamefont  [1]{#1}%
\providecommand \bibfnamefont [1]{#1}%
\providecommand \citenamefont [1]{#1}%
\providecommand \href@noop [0]{\@secondoftwo}%
\providecommand \href [0]{\begingroup \@sanitize@url \@href}%
\providecommand \@href[1]{\@@startlink{#1}\@@href}%
\providecommand \@@href[1]{\endgroup#1\@@endlink}%
\providecommand \@sanitize@url [0]{\catcode `\\12\catcode `\$12\catcode
  `\&12\catcode `\#12\catcode `\^12\catcode `\_12\catcode `\%12\relax}%
\providecommand \@@startlink[1]{}%
\providecommand \@@endlink[0]{}%
\providecommand \url  [0]{\begingroup\@sanitize@url \@url }%
\providecommand \@url [1]{\endgroup\@href {#1}{\urlprefix }}%
\providecommand \urlprefix  [0]{URL }%
\providecommand \Eprint [0]{\href }%
\providecommand \doibase [0]{http://dx.doi.org/}%
\providecommand \selectlanguage [0]{\@gobble}%
\providecommand \bibinfo  [0]{\@secondoftwo}%
\providecommand \bibfield  [0]{\@secondoftwo}%
\providecommand \translation [1]{[#1]}%
\providecommand \BibitemOpen [0]{}%
\providecommand \bibitemStop [0]{}%
\providecommand \bibitemNoStop [0]{.\EOS\space}%
\providecommand \EOS [0]{\spacefactor3000\relax}%
\providecommand \BibitemShut  [1]{\csname bibitem#1\endcsname}%
\let\auto@bib@innerbib\@empty
\bibitem [{\citenamefont {Heinrich}\ \emph {et~al.}(1987)\citenamefont
  {Heinrich}, \citenamefont {Urquhart}, \citenamefont {Arrott}, \citenamefont
  {Cochran}, \citenamefont {Myrtle},\ and\ \citenamefont
  {Purcell}}]{heinrichFerromagneticresonanceStudyUltrathin1987}%
  \BibitemOpen
  \bibfield  {author} {\bibinfo {author} {\bibfnamefont {B.}~\bibnamefont
  {Heinrich}}, \bibinfo {author} {\bibfnamefont {K.~B.}\ \bibnamefont
  {Urquhart}}, \bibinfo {author} {\bibfnamefont {A.~S.}\ \bibnamefont
  {Arrott}}, \bibinfo {author} {\bibfnamefont {J.~F.}\ \bibnamefont {Cochran}},
  \bibinfo {author} {\bibfnamefont {K.}~\bibnamefont {Myrtle}}, \ and\ \bibinfo
  {author} {\bibfnamefont {S.~T.}\ \bibnamefont {Purcell}},\ }\href {\doibase
  10.1103/PhysRevLett.59.1756} {\bibfield  {journal} {\bibinfo  {journal}
  {Phys. Rev. Lett.}\ }\textbf {\bibinfo {volume} {59}},\ \bibinfo {pages}
  {1756} (\bibinfo {year} {1987})}\BibitemShut {NoStop}%
\bibitem [{\citenamefont {Celinski}\ and\ \citenamefont
  {Heinrich}(1991)}]{celinskiFerromagneticResonanceLinewidth1991}%
  \BibitemOpen
  \bibfield  {author} {\bibinfo {author} {\bibfnamefont {Z.}~\bibnamefont
  {Celinski}}\ and\ \bibinfo {author} {\bibfnamefont {B.}~\bibnamefont
  {Heinrich}},\ }\href {\doibase 10.1063/1.350110} {\bibfield  {journal}
  {\bibinfo  {journal} {Journal of Applied Physics}\ }\textbf {\bibinfo
  {volume} {70}},\ \bibinfo {pages} {5935} (\bibinfo {year}
  {1991})}\BibitemShut {NoStop}%
\bibitem [{\citenamefont {Mizukami}\ \emph
  {et~al.}(2001{\natexlab{a}})\citenamefont {Mizukami}, \citenamefont {Ando},\
  and\ \citenamefont {Miyazaki}}]{mizukamiFerromagneticResonanceLinewidth2001}%
  \BibitemOpen
  \bibfield  {author} {\bibinfo {author} {\bibfnamefont {S.}~\bibnamefont
  {Mizukami}}, \bibinfo {author} {\bibfnamefont {Y.}~\bibnamefont {Ando}}, \
  and\ \bibinfo {author} {\bibfnamefont {T.}~\bibnamefont {Miyazaki}},\
  }\href@noop {} {\bibfield  {journal} {\bibinfo  {journal} {Journal of
  Magnetism and Magnetic Materials}\ ,\ \bibinfo {pages} {3}} (\bibinfo {year}
  {2001}{\natexlab{a}})}\BibitemShut {NoStop}%
\bibitem [{\citenamefont {Mizukami}\ \emph
  {et~al.}(2001{\natexlab{b}})\citenamefont {Mizukami}, \citenamefont {Ando},\
  and\ \citenamefont {Miyazaki}}]{mizukamiStudyFerromagneticResonance2001}%
  \BibitemOpen
  \bibfield  {author} {\bibinfo {author} {\bibfnamefont {S.}~\bibnamefont
  {Mizukami}}, \bibinfo {author} {\bibfnamefont {Y.}~\bibnamefont {Ando}}, \
  and\ \bibinfo {author} {\bibfnamefont {T.}~\bibnamefont {Miyazaki}},\ }\href
  {\doibase 10.1143/JJAP.40.580} {\bibfield  {journal} {\bibinfo  {journal}
  {Jpn. J. Appl. Phys.}\ }\textbf {\bibinfo {volume} {40}},\ \bibinfo {pages}
  {580} (\bibinfo {year} {2001}{\natexlab{b}})}\BibitemShut {NoStop}%
\bibitem [{\citenamefont {Mizukami}\ \emph {et~al.}(2002)\citenamefont
  {Mizukami}, \citenamefont {Ando},\ and\ \citenamefont
  {Miyazaki}}]{mizukamiEffectSpinDiffusion2002}%
  \BibitemOpen
  \bibfield  {author} {\bibinfo {author} {\bibfnamefont {S.}~\bibnamefont
  {Mizukami}}, \bibinfo {author} {\bibfnamefont {Y.}~\bibnamefont {Ando}}, \
  and\ \bibinfo {author} {\bibfnamefont {T.}~\bibnamefont {Miyazaki}},\ }\href
  {\doibase 10.1103/PhysRevB.66.104413} {\bibfield  {journal} {\bibinfo
  {journal} {Phys. Rev. B}\ }\textbf {\bibinfo {volume} {66}},\ \bibinfo
  {pages} {104413} (\bibinfo {year} {2002})}\BibitemShut {NoStop}%
\bibitem [{\citenamefont {Ingvarsson}\ \emph {et~al.}(2002)\citenamefont
  {Ingvarsson}, \citenamefont {Ritchie}, \citenamefont {Liu}, \citenamefont
  {Xiao}, \citenamefont {Slonczewski}, \citenamefont {Trouilloud},\ and\
  \citenamefont {Koch}}]{ingvarssonRoleElectronScattering2002}%
  \BibitemOpen
  \bibfield  {author} {\bibinfo {author} {\bibfnamefont {S.}~\bibnamefont
  {Ingvarsson}}, \bibinfo {author} {\bibfnamefont {L.}~\bibnamefont {Ritchie}},
  \bibinfo {author} {\bibfnamefont {X.~Y.}\ \bibnamefont {Liu}}, \bibinfo
  {author} {\bibfnamefont {G.}~\bibnamefont {Xiao}}, \bibinfo {author}
  {\bibfnamefont {J.~C.}\ \bibnamefont {Slonczewski}}, \bibinfo {author}
  {\bibfnamefont {P.~L.}\ \bibnamefont {Trouilloud}}, \ and\ \bibinfo {author}
  {\bibfnamefont {R.~H.}\ \bibnamefont {Koch}},\ }\href {\doibase
  10.1103/PhysRevB.66.214416} {\bibfield  {journal} {\bibinfo  {journal} {Phys.
  Rev. B}\ }\textbf {\bibinfo {volume} {66}},\ \bibinfo {pages} {214416}
  (\bibinfo {year} {2002})}\BibitemShut {NoStop}%
\bibitem [{\citenamefont {Lubitz}\ \emph {et~al.}(2003)\citenamefont {Lubitz},
  \citenamefont {Cheng},\ and\ \citenamefont
  {Rachford}}]{lubitzIncreaseMagneticDamping2003}%
  \BibitemOpen
  \bibfield  {author} {\bibinfo {author} {\bibfnamefont {P.}~\bibnamefont
  {Lubitz}}, \bibinfo {author} {\bibfnamefont {S.~F.}\ \bibnamefont {Cheng}}, \
  and\ \bibinfo {author} {\bibfnamefont {F.~J.}\ \bibnamefont {Rachford}},\
  }\href {\doibase 10.1063/1.1557340} {\bibfield  {journal} {\bibinfo
  {journal} {Journal of Applied Physics}\ }\textbf {\bibinfo {volume} {93}},\
  \bibinfo {pages} {8283} (\bibinfo {year} {2003})}\BibitemShut {NoStop}%
\bibitem [{\citenamefont
  {Sarma}(2004)}]{sarmaSpintronicsFundamentalsApplications2004}%
  \BibitemOpen
  \bibfield  {author} {\bibinfo {author} {\bibfnamefont {S.~D.}\ \bibnamefont
  {Sarma}},\ }\href@noop {} {\bibfield  {journal} {\bibinfo  {journal} {Rev.
  Mod. Phys.}\ }\textbf {\bibinfo {volume} {76}},\ \bibinfo {pages} {88}
  (\bibinfo {year} {2004})}\BibitemShut {NoStop}%
\bibitem [{\citenamefont {Tserkovnyak}\ \emph {et~al.}(2005)\citenamefont
  {Tserkovnyak}, \citenamefont {Brataas}, \citenamefont {Bauer},\ and\
  \citenamefont {Halperin}}]{tserkovnyakNonlocalMagnetizationDynamics2005}%
  \BibitemOpen
  \bibfield  {author} {\bibinfo {author} {\bibfnamefont {Y.}~\bibnamefont
  {Tserkovnyak}}, \bibinfo {author} {\bibfnamefont {A.}~\bibnamefont
  {Brataas}}, \bibinfo {author} {\bibfnamefont {G.~E.~W.}\ \bibnamefont
  {Bauer}}, \ and\ \bibinfo {author} {\bibfnamefont {B.~I.}\ \bibnamefont
  {Halperin}},\ }\href@noop {} {\bibfield  {journal} {\bibinfo  {journal} {Rev.
  Mod. Phys.}\ }\textbf {\bibinfo {volume} {77}},\ \bibinfo {pages} {47}
  (\bibinfo {year} {2005})}\BibitemShut {NoStop}%
\bibitem [{\citenamefont {Azevedo}\ \emph {et~al.}(2005)\citenamefont
  {Azevedo}, \citenamefont {Vilela~Le{\~a}o}, \citenamefont
  {{Rodriguez-Suarez}}, \citenamefont {Oliveira},\ and\ \citenamefont
  {Rezende}}]{azevedoDcEffectFerromagnetic2005}%
  \BibitemOpen
  \bibfield  {author} {\bibinfo {author} {\bibfnamefont {A.}~\bibnamefont
  {Azevedo}}, \bibinfo {author} {\bibfnamefont {L.~H.}\ \bibnamefont
  {Vilela~Le{\~a}o}}, \bibinfo {author} {\bibfnamefont {R.~L.}\ \bibnamefont
  {{Rodriguez-Suarez}}}, \bibinfo {author} {\bibfnamefont {A.~B.}\ \bibnamefont
  {Oliveira}}, \ and\ \bibinfo {author} {\bibfnamefont {S.~M.}\ \bibnamefont
  {Rezende}},\ }\href {\doibase 10.1063/1.1855251} {\bibfield  {journal}
  {\bibinfo  {journal} {Journal of Applied Physics}\ }\textbf {\bibinfo
  {volume} {97}},\ \bibinfo {pages} {10C715} (\bibinfo {year}
  {2005})}\BibitemShut {NoStop}%
\bibitem [{\citenamefont {Saitoh}\ \emph {et~al.}(2006)\citenamefont {Saitoh},
  \citenamefont {Ueda}, \citenamefont {Miyajima},\ and\ \citenamefont
  {Tatara}}]{saitohConversionSpinCurrent2006}%
  \BibitemOpen
  \bibfield  {author} {\bibinfo {author} {\bibfnamefont {E.}~\bibnamefont
  {Saitoh}}, \bibinfo {author} {\bibfnamefont {M.}~\bibnamefont {Ueda}},
  \bibinfo {author} {\bibfnamefont {H.}~\bibnamefont {Miyajima}}, \ and\
  \bibinfo {author} {\bibfnamefont {G.}~\bibnamefont {Tatara}},\ }\href
  {\doibase 10.1063/1.2199473} {\bibfield  {journal} {\bibinfo  {journal}
  {Appl. Phys. Lett.}\ }\textbf {\bibinfo {volume} {88}},\ \bibinfo {pages}
  {182509} (\bibinfo {year} {2006})}\BibitemShut {NoStop}%
\bibitem [{\citenamefont {Ando}\ \emph {et~al.}(2008)\citenamefont {Ando},
  \citenamefont {Takahashi}, \citenamefont {Harii}, \citenamefont {Sasage},
  \citenamefont {Ieda}, \citenamefont {Maekawa},\ and\ \citenamefont
  {Saitoh}}]{andoElectricManipulationSpin2008}%
  \BibitemOpen
  \bibfield  {author} {\bibinfo {author} {\bibfnamefont {K.}~\bibnamefont
  {Ando}}, \bibinfo {author} {\bibfnamefont {S.}~\bibnamefont {Takahashi}},
  \bibinfo {author} {\bibfnamefont {K.}~\bibnamefont {Harii}}, \bibinfo
  {author} {\bibfnamefont {K.}~\bibnamefont {Sasage}}, \bibinfo {author}
  {\bibfnamefont {J.}~\bibnamefont {Ieda}}, \bibinfo {author} {\bibfnamefont
  {S.}~\bibnamefont {Maekawa}}, \ and\ \bibinfo {author} {\bibfnamefont
  {E.}~\bibnamefont {Saitoh}},\ }\href {\doibase
  10.1103/PhysRevLett.101.036601} {\bibfield  {journal} {\bibinfo  {journal}
  {Phys. Rev. Lett.}\ }\textbf {\bibinfo {volume} {101}},\ \bibinfo {pages}
  {036601} (\bibinfo {year} {2008})}\BibitemShut {NoStop}%
\bibitem [{\citenamefont {Ando}\ \emph {et~al.}(2009)\citenamefont {Ando},
  \citenamefont {Ieda}, \citenamefont {Sasage}, \citenamefont {Takahashi},
  \citenamefont {Maekawa},\ and\ \citenamefont
  {Saitoh}}]{andoElectricDetectionSpin2009}%
  \BibitemOpen
  \bibfield  {author} {\bibinfo {author} {\bibfnamefont {K.}~\bibnamefont
  {Ando}}, \bibinfo {author} {\bibfnamefont {J.}~\bibnamefont {Ieda}}, \bibinfo
  {author} {\bibfnamefont {K.}~\bibnamefont {Sasage}}, \bibinfo {author}
  {\bibfnamefont {S.}~\bibnamefont {Takahashi}}, \bibinfo {author}
  {\bibfnamefont {S.}~\bibnamefont {Maekawa}}, \ and\ \bibinfo {author}
  {\bibfnamefont {E.}~\bibnamefont {Saitoh}},\ }\href {\doibase
  10.1063/1.3167826} {\bibfield  {journal} {\bibinfo  {journal} {Appl. Phys.
  Lett.}\ }\textbf {\bibinfo {volume} {94}},\ \bibinfo {pages} {262505}
  (\bibinfo {year} {2009})}\BibitemShut {NoStop}%
\bibitem [{\citenamefont {Ando}\ \emph {et~al.}(2011)\citenamefont {Ando},
  \citenamefont {Takahashi}, \citenamefont {Ieda}, \citenamefont {Kurebayashi},
  \citenamefont {Trypiniotis}, \citenamefont {Barnes}, \citenamefont
  {Maekawa},\ and\ \citenamefont {Saitoh}}]{andoElectricallySpin2011}%
  \BibitemOpen
  \bibfield  {author} {\bibinfo {author} {\bibfnamefont {K.}~\bibnamefont
  {Ando}}, \bibinfo {author} {\bibfnamefont {S.}~\bibnamefont {Takahashi}},
  \bibinfo {author} {\bibfnamefont {J.}~\bibnamefont {Ieda}}, \bibinfo {author}
  {\bibfnamefont {H.}~\bibnamefont {Kurebayashi}}, \bibinfo {author}
  {\bibfnamefont {T.}~\bibnamefont {Trypiniotis}}, \bibinfo {author}
  {\bibfnamefont {C.~H.~W.}\ \bibnamefont {Barnes}}, \bibinfo {author}
  {\bibfnamefont {S.}~\bibnamefont {Maekawa}}, \ and\ \bibinfo {author}
  {\bibfnamefont {E.}~\bibnamefont {Saitoh}},\ }\href {\doibase
  10.1038/nmat3052} {\bibfield  {journal} {\bibinfo  {journal} {Nature Mater}\
  }\textbf {\bibinfo {volume} {10}},\ \bibinfo {pages} {655} (\bibinfo {year}
  {2011})}\BibitemShut {NoStop}%
\bibitem [{\citenamefont {Mosendz}\ \emph
  {et~al.}(2010{\natexlab{a}})\citenamefont {Mosendz}, \citenamefont
  {Vlaminck}, \citenamefont {Pearson}, \citenamefont {Fradin}, \citenamefont
  {Bauer}, \citenamefont {Bader},\ and\ \citenamefont
  {Hoffmann}}]{mosendzDetectionQuantificationInverse2010}%
  \BibitemOpen
  \bibfield  {author} {\bibinfo {author} {\bibfnamefont {O.}~\bibnamefont
  {Mosendz}}, \bibinfo {author} {\bibfnamefont {V.}~\bibnamefont {Vlaminck}},
  \bibinfo {author} {\bibfnamefont {J.~E.}\ \bibnamefont {Pearson}}, \bibinfo
  {author} {\bibfnamefont {F.~Y.}\ \bibnamefont {Fradin}}, \bibinfo {author}
  {\bibfnamefont {G.~E.~W.}\ \bibnamefont {Bauer}}, \bibinfo {author}
  {\bibfnamefont {S.~D.}\ \bibnamefont {Bader}}, \ and\ \bibinfo {author}
  {\bibfnamefont {A.}~\bibnamefont {Hoffmann}},\ }\href {\doibase
  10.1103/PhysRevB.82.214403} {\bibfield  {journal} {\bibinfo  {journal} {Phys.
  Rev. B}\ }\textbf {\bibinfo {volume} {82}},\ \bibinfo {pages} {214403}
  (\bibinfo {year} {2010}{\natexlab{a}})}\BibitemShut {NoStop}%
\bibitem [{\citenamefont {Mosendz}\ \emph
  {et~al.}(2010{\natexlab{b}})\citenamefont {Mosendz}, \citenamefont {Pearson},
  \citenamefont {Fradin}, \citenamefont {Bauer}, \citenamefont {Bader},\ and\
  \citenamefont {Hoffmann}}]{mosendzQuantifyingSpinHall2010}%
  \BibitemOpen
  \bibfield  {author} {\bibinfo {author} {\bibfnamefont {O.}~\bibnamefont
  {Mosendz}}, \bibinfo {author} {\bibfnamefont {J.~E.}\ \bibnamefont
  {Pearson}}, \bibinfo {author} {\bibfnamefont {F.~Y.}\ \bibnamefont {Fradin}},
  \bibinfo {author} {\bibfnamefont {G.~E.~W.}\ \bibnamefont {Bauer}}, \bibinfo
  {author} {\bibfnamefont {S.~D.}\ \bibnamefont {Bader}}, \ and\ \bibinfo
  {author} {\bibfnamefont {A.}~\bibnamefont {Hoffmann}},\ }\href {\doibase
  10.1103/PhysRevLett.104.046601} {\bibfield  {journal} {\bibinfo  {journal}
  {Phys. Rev. Lett.}\ }\textbf {\bibinfo {volume} {104}},\ \bibinfo {pages}
  {046601} (\bibinfo {year} {2010}{\natexlab{b}})}\BibitemShut {NoStop}%
\bibitem [{\citenamefont {Czeschka}\ \emph {et~al.}(2011)\citenamefont
  {Czeschka}, \citenamefont {Dreher}, \citenamefont {Brandt}, \citenamefont
  {Weiler}, \citenamefont {Althammer}, \citenamefont {Imort}, \citenamefont
  {Reiss}, \citenamefont {Thomas}, \citenamefont {Schoch}, \citenamefont
  {Limmer}, \citenamefont {Huebl}, \citenamefont {Gross},\ and\ \citenamefont
  {Goennenwein}}]{czeschkaScalingBehaviorSpin2011}%
  \BibitemOpen
  \bibfield  {author} {\bibinfo {author} {\bibfnamefont {F.~D.}\ \bibnamefont
  {Czeschka}}, \bibinfo {author} {\bibfnamefont {L.}~\bibnamefont {Dreher}},
  \bibinfo {author} {\bibfnamefont {M.~S.}\ \bibnamefont {Brandt}}, \bibinfo
  {author} {\bibfnamefont {M.}~\bibnamefont {Weiler}}, \bibinfo {author}
  {\bibfnamefont {M.}~\bibnamefont {Althammer}}, \bibinfo {author}
  {\bibfnamefont {I.-M.}\ \bibnamefont {Imort}}, \bibinfo {author}
  {\bibfnamefont {G.}~\bibnamefont {Reiss}}, \bibinfo {author} {\bibfnamefont
  {A.}~\bibnamefont {Thomas}}, \bibinfo {author} {\bibfnamefont
  {W.}~\bibnamefont {Schoch}}, \bibinfo {author} {\bibfnamefont
  {W.}~\bibnamefont {Limmer}}, \bibinfo {author} {\bibfnamefont
  {H.}~\bibnamefont {Huebl}}, \bibinfo {author} {\bibfnamefont
  {R.}~\bibnamefont {Gross}}, \ and\ \bibinfo {author} {\bibfnamefont
  {S.~T.~B.}\ \bibnamefont {Goennenwein}},\ }\href {\doibase
  10.1103/PhysRevLett.107.046601} {\bibfield  {journal} {\bibinfo  {journal}
  {Phys. Rev. Lett.}\ }\textbf {\bibinfo {volume} {107}},\ \bibinfo {pages}
  {046601} (\bibinfo {year} {2011})}\BibitemShut {NoStop}%
\bibitem [{\citenamefont {Miron}\ \emph {et~al.}(2011)\citenamefont {Miron},
  \citenamefont {Garello}, \citenamefont {Gaudin}, \citenamefont {Zermatten},
  \citenamefont {Costache}, \citenamefont {Auffret}, \citenamefont {Bandiera},
  \citenamefont {Rodmacq}, \citenamefont {Schuhl},\ and\ \citenamefont
  {Gambardella}}]{mironPerpendicularSwitchingSingle2011}%
  \BibitemOpen
  \bibfield  {author} {\bibinfo {author} {\bibfnamefont {I.~M.}\ \bibnamefont
  {Miron}}, \bibinfo {author} {\bibfnamefont {K.}~\bibnamefont {Garello}},
  \bibinfo {author} {\bibfnamefont {G.}~\bibnamefont {Gaudin}}, \bibinfo
  {author} {\bibfnamefont {P.-J.}\ \bibnamefont {Zermatten}}, \bibinfo {author}
  {\bibfnamefont {M.~V.}\ \bibnamefont {Costache}}, \bibinfo {author}
  {\bibfnamefont {S.}~\bibnamefont {Auffret}}, \bibinfo {author} {\bibfnamefont
  {S.}~\bibnamefont {Bandiera}}, \bibinfo {author} {\bibfnamefont
  {B.}~\bibnamefont {Rodmacq}}, \bibinfo {author} {\bibfnamefont
  {A.}~\bibnamefont {Schuhl}}, \ and\ \bibinfo {author} {\bibfnamefont
  {P.}~\bibnamefont {Gambardella}},\ }\href {\doibase 10.1038/nature10309}
  {\bibfield  {journal} {\bibinfo  {journal} {Nature}\ }\textbf {\bibinfo
  {volume} {476}},\ \bibinfo {pages} {189} (\bibinfo {year}
  {2011})}\BibitemShut {NoStop}%
\bibitem [{\citenamefont {Liu}\ \emph {et~al.}(2011)\citenamefont {Liu},
  \citenamefont {Moriyama}, \citenamefont {Ralph},\ and\ \citenamefont
  {Buhrman}}]{liuSpinTorqueFerromagneticResonance2011}%
  \BibitemOpen
  \bibfield  {author} {\bibinfo {author} {\bibfnamefont {L.}~\bibnamefont
  {Liu}}, \bibinfo {author} {\bibfnamefont {T.}~\bibnamefont {Moriyama}},
  \bibinfo {author} {\bibfnamefont {D.~C.}\ \bibnamefont {Ralph}}, \ and\
  \bibinfo {author} {\bibfnamefont {R.~A.}\ \bibnamefont {Buhrman}},\ }\href
  {\doibase 10.1103/PhysRevLett.106.036601} {\bibfield  {journal} {\bibinfo
  {journal} {Phys. Rev. Lett.}\ }\textbf {\bibinfo {volume} {106}},\ \bibinfo
  {pages} {036601} (\bibinfo {year} {2011})}\BibitemShut {NoStop}%
\bibitem [{\citenamefont {Liu}\ \emph {et~al.}(2012)\citenamefont {Liu},
  \citenamefont {Pai}, \citenamefont {Li}, \citenamefont {Tseng}, \citenamefont
  {Ralph},\ and\ \citenamefont {Buhrman}}]{liuSpinTorqueSwitchingGiant2012}%
  \BibitemOpen
  \bibfield  {author} {\bibinfo {author} {\bibfnamefont {L.}~\bibnamefont
  {Liu}}, \bibinfo {author} {\bibfnamefont {C.-F.}\ \bibnamefont {Pai}},
  \bibinfo {author} {\bibfnamefont {Y.}~\bibnamefont {Li}}, \bibinfo {author}
  {\bibfnamefont {H.~W.}\ \bibnamefont {Tseng}}, \bibinfo {author}
  {\bibfnamefont {D.~C.}\ \bibnamefont {Ralph}}, \ and\ \bibinfo {author}
  {\bibfnamefont {R.~A.}\ \bibnamefont {Buhrman}},\ }\href {\doibase
  10.1126/science.1218197} {\bibfield  {journal} {\bibinfo  {journal}
  {Science}\ }\textbf {\bibinfo {volume} {336}},\ \bibinfo {pages} {555}
  (\bibinfo {year} {2012})}\BibitemShut {NoStop}%
\bibitem [{\citenamefont {Kajiwara}\ \emph {et~al.}(2010)\citenamefont
  {Kajiwara}, \citenamefont {Harii}, \citenamefont {Takahashi}, \citenamefont
  {Ohe}, \citenamefont {Uchida}, \citenamefont {Mizuguchi}, \citenamefont
  {Umezawa}, \citenamefont {Kawai}, \citenamefont {Ando}, \citenamefont
  {Takanashi}, \citenamefont {Maekawa},\ and\ \citenamefont
  {Saitoh}}]{kajiwaraTransmissionElectricalSignals2010}%
  \BibitemOpen
  \bibfield  {author} {\bibinfo {author} {\bibfnamefont {Y.}~\bibnamefont
  {Kajiwara}}, \bibinfo {author} {\bibfnamefont {K.}~\bibnamefont {Harii}},
  \bibinfo {author} {\bibfnamefont {S.}~\bibnamefont {Takahashi}}, \bibinfo
  {author} {\bibfnamefont {J.}~\bibnamefont {Ohe}}, \bibinfo {author}
  {\bibfnamefont {K.}~\bibnamefont {Uchida}}, \bibinfo {author} {\bibfnamefont
  {M.}~\bibnamefont {Mizuguchi}}, \bibinfo {author} {\bibfnamefont
  {H.}~\bibnamefont {Umezawa}}, \bibinfo {author} {\bibfnamefont
  {H.}~\bibnamefont {Kawai}}, \bibinfo {author} {\bibfnamefont
  {K.}~\bibnamefont {Ando}}, \bibinfo {author} {\bibfnamefont {K.}~\bibnamefont
  {Takanashi}}, \bibinfo {author} {\bibfnamefont {S.}~\bibnamefont {Maekawa}},
  \ and\ \bibinfo {author} {\bibfnamefont {E.}~\bibnamefont {Saitoh}},\ }\href
  {\doibase 10.1038/nature08876} {\bibfield  {journal} {\bibinfo  {journal}
  {Nature}\ }\textbf {\bibinfo {volume} {464}},\ \bibinfo {pages} {262}
  (\bibinfo {year} {2010})}\BibitemShut {NoStop}%
\bibitem [{\citenamefont {Bai}\ \emph {et~al.}(2013)\citenamefont {Bai},
  \citenamefont {Hyde}, \citenamefont {Gui}, \citenamefont {Hu}, \citenamefont
  {Vlaminck}, \citenamefont {Pearson}, \citenamefont {Bader},\ and\
  \citenamefont {Hoffmann}}]{baiUniversalMethodSeparating2013}%
  \BibitemOpen
  \bibfield  {author} {\bibinfo {author} {\bibfnamefont {L.}~\bibnamefont
  {Bai}}, \bibinfo {author} {\bibfnamefont {P.}~\bibnamefont {Hyde}}, \bibinfo
  {author} {\bibfnamefont {Y.~S.}\ \bibnamefont {Gui}}, \bibinfo {author}
  {\bibfnamefont {C.-M.}\ \bibnamefont {Hu}}, \bibinfo {author} {\bibfnamefont
  {V.}~\bibnamefont {Vlaminck}}, \bibinfo {author} {\bibfnamefont {J.~E.}\
  \bibnamefont {Pearson}}, \bibinfo {author} {\bibfnamefont {S.~D.}\
  \bibnamefont {Bader}}, \ and\ \bibinfo {author} {\bibfnamefont
  {A.}~\bibnamefont {Hoffmann}},\ }\href {\doibase
  10.1103/PhysRevLett.111.217602} {\bibfield  {journal} {\bibinfo  {journal}
  {Phys. Rev. Lett.}\ }\textbf {\bibinfo {volume} {111}},\ \bibinfo {pages}
  {217602} (\bibinfo {year} {2013})}\BibitemShut {NoStop}%
\bibitem [{\citenamefont {Sandweg}\ \emph {et~al.}(2011)\citenamefont
  {Sandweg}, \citenamefont {Kajiwara}, \citenamefont {Chumak}, \citenamefont
  {Serga}, \citenamefont {Vasyuchka}, \citenamefont {Jungfleisch},
  \citenamefont {Saitoh},\ and\ \citenamefont
  {Hillebrands}}]{sandwegSpinPumpingParametrically2011}%
  \BibitemOpen
  \bibfield  {author} {\bibinfo {author} {\bibfnamefont {C.~W.}\ \bibnamefont
  {Sandweg}}, \bibinfo {author} {\bibfnamefont {Y.}~\bibnamefont {Kajiwara}},
  \bibinfo {author} {\bibfnamefont {A.~V.}\ \bibnamefont {Chumak}}, \bibinfo
  {author} {\bibfnamefont {A.~A.}\ \bibnamefont {Serga}}, \bibinfo {author}
  {\bibfnamefont {V.~I.}\ \bibnamefont {Vasyuchka}}, \bibinfo {author}
  {\bibfnamefont {M.~B.}\ \bibnamefont {Jungfleisch}}, \bibinfo {author}
  {\bibfnamefont {E.}~\bibnamefont {Saitoh}}, \ and\ \bibinfo {author}
  {\bibfnamefont {B.}~\bibnamefont {Hillebrands}},\ }\href {\doibase
  10.1103/PhysRevLett.106.216601} {\bibfield  {journal} {\bibinfo  {journal}
  {Phys. Rev. Lett.}\ }\textbf {\bibinfo {volume} {106}},\ \bibinfo {pages}
  {216601} (\bibinfo {year} {2011})}\BibitemShut {NoStop}%
\bibitem [{\citenamefont {Sinova}\ \emph {et~al.}(2015)\citenamefont {Sinova},
  \citenamefont {Valenzuela}, \citenamefont {Wunderlich}, \citenamefont
  {Back},\ and\ \citenamefont {Jungwirth}}]{sinovaSpinHallEffects2015}%
  \BibitemOpen
  \bibfield  {author} {\bibinfo {author} {\bibfnamefont {J.}~\bibnamefont
  {Sinova}}, \bibinfo {author} {\bibfnamefont {S.~O.}\ \bibnamefont
  {Valenzuela}}, \bibinfo {author} {\bibfnamefont {J.}~\bibnamefont
  {Wunderlich}}, \bibinfo {author} {\bibfnamefont {C.~H.}\ \bibnamefont
  {Back}}, \ and\ \bibinfo {author} {\bibfnamefont {T.}~\bibnamefont
  {Jungwirth}},\ }\href {\doibase 10.1103/RevModPhys.87.1213} {\bibfield
  {journal} {\bibinfo  {journal} {Rev. Mod. Phys.}\ }\textbf {\bibinfo {volume}
  {87}},\ \bibinfo {pages} {1213} (\bibinfo {year} {2015})}\BibitemShut
  {NoStop}%
\bibitem [{\citenamefont {Berger}(1996)}]{bergerEmissionSpinWaves1996}%
  \BibitemOpen
  \bibfield  {author} {\bibinfo {author} {\bibfnamefont {L.}~\bibnamefont
  {Berger}},\ }\href {\doibase 10.1103/PhysRevB.54.9353} {\bibfield  {journal}
  {\bibinfo  {journal} {Phys. Rev. B}\ }\textbf {\bibinfo {volume} {54}},\
  \bibinfo {pages} {9353} (\bibinfo {year} {1996})}\BibitemShut {NoStop}%
\bibitem [{\citenamefont {Hellman}\ \emph {et~al.}(2017)\citenamefont
  {Hellman}, \citenamefont {Hoffmann}, \citenamefont {Tserkovnyak},
  \citenamefont {Beach}, \citenamefont {Fullerton}, \citenamefont {Leighton},
  \citenamefont {MacDonald}, \citenamefont {Ralph}, \citenamefont {Arena},
  \citenamefont {D{\"u}rr}, \citenamefont {Fischer}, \citenamefont {Grollier},
  \citenamefont {Heremans}, \citenamefont {Jungwirth}, \citenamefont {Kimel},
  \citenamefont {Koopmans}, \citenamefont {Krivorotov}, \citenamefont {May},
  \citenamefont {{Petford-Long}}, \citenamefont {Rondinelli}, \citenamefont
  {Samarth}, \citenamefont {Schuller}, \citenamefont {Slavin}, \citenamefont
  {Stiles}, \citenamefont {Tchernyshyov}, \citenamefont {Thiaville},\ and\
  \citenamefont {Zink}}]{hellmanInterfaceinducedPhenomenaMagnetism2017}%
  \BibitemOpen
  \bibfield  {author} {\bibinfo {author} {\bibfnamefont {F.}~\bibnamefont
  {Hellman}}, \bibinfo {author} {\bibfnamefont {A.}~\bibnamefont {Hoffmann}},
  \bibinfo {author} {\bibfnamefont {Y.}~\bibnamefont {Tserkovnyak}}, \bibinfo
  {author} {\bibfnamefont {G.~S.~D.}\ \bibnamefont {Beach}}, \bibinfo {author}
  {\bibfnamefont {E.~E.}\ \bibnamefont {Fullerton}}, \bibinfo {author}
  {\bibfnamefont {C.}~\bibnamefont {Leighton}}, \bibinfo {author}
  {\bibfnamefont {A.~H.}\ \bibnamefont {MacDonald}}, \bibinfo {author}
  {\bibfnamefont {D.~C.}\ \bibnamefont {Ralph}}, \bibinfo {author}
  {\bibfnamefont {D.~A.}\ \bibnamefont {Arena}}, \bibinfo {author}
  {\bibfnamefont {H.~A.}\ \bibnamefont {D{\"u}rr}}, \bibinfo {author}
  {\bibfnamefont {P.}~\bibnamefont {Fischer}}, \bibinfo {author} {\bibfnamefont
  {J.}~\bibnamefont {Grollier}}, \bibinfo {author} {\bibfnamefont {J.~P.}\
  \bibnamefont {Heremans}}, \bibinfo {author} {\bibfnamefont {T.}~\bibnamefont
  {Jungwirth}}, \bibinfo {author} {\bibfnamefont {A.~V.}\ \bibnamefont
  {Kimel}}, \bibinfo {author} {\bibfnamefont {B.}~\bibnamefont {Koopmans}},
  \bibinfo {author} {\bibfnamefont {I.~N.}\ \bibnamefont {Krivorotov}},
  \bibinfo {author} {\bibfnamefont {S.~J.}\ \bibnamefont {May}}, \bibinfo
  {author} {\bibfnamefont {A.~K.}\ \bibnamefont {{Petford-Long}}}, \bibinfo
  {author} {\bibfnamefont {J.~M.}\ \bibnamefont {Rondinelli}}, \bibinfo
  {author} {\bibfnamefont {N.}~\bibnamefont {Samarth}}, \bibinfo {author}
  {\bibfnamefont {I.~K.}\ \bibnamefont {Schuller}}, \bibinfo {author}
  {\bibfnamefont {A.~N.}\ \bibnamefont {Slavin}}, \bibinfo {author}
  {\bibfnamefont {M.~D.}\ \bibnamefont {Stiles}}, \bibinfo {author}
  {\bibfnamefont {O.}~\bibnamefont {Tchernyshyov}}, \bibinfo {author}
  {\bibfnamefont {A.}~\bibnamefont {Thiaville}}, \ and\ \bibinfo {author}
  {\bibfnamefont {B.~L.}\ \bibnamefont {Zink}},\ }\href {\doibase
  10.1103/RevModPhys.89.025006} {\bibfield  {journal} {\bibinfo  {journal}
  {Rev. Mod. Phys.}\ }\textbf {\bibinfo {volume} {89}},\ \bibinfo {pages}
  {025006} (\bibinfo {year} {2017})}\BibitemShut {NoStop}%
\bibitem [{\citenamefont {Tserkovnyak}\ \emph
  {et~al.}(2002{\natexlab{a}})\citenamefont {Tserkovnyak}, \citenamefont
  {Brataas},\ and\ \citenamefont
  {Bauer}}]{tserkovnyakSpinPumpingMagnetization2002}%
  \BibitemOpen
  \bibfield  {author} {\bibinfo {author} {\bibfnamefont {Y.}~\bibnamefont
  {Tserkovnyak}}, \bibinfo {author} {\bibfnamefont {A.}~\bibnamefont
  {Brataas}}, \ and\ \bibinfo {author} {\bibfnamefont {G.~E.~W.}\ \bibnamefont
  {Bauer}},\ }\href {\doibase 10.1103/PhysRevB.66.224403} {\bibfield  {journal}
  {\bibinfo  {journal} {Phys. Rev. B}\ }\textbf {\bibinfo {volume} {66}},\
  \bibinfo {pages} {224403} (\bibinfo {year} {2002}{\natexlab{a}})}\BibitemShut
  {NoStop}%
\bibitem [{\citenamefont {Tserkovnyak}\ \emph
  {et~al.}(2002{\natexlab{b}})\citenamefont {Tserkovnyak}, \citenamefont
  {Brataas},\ and\ \citenamefont
  {Bauer}}]{tserkovnyakEnhancedGilbertDamping2002}%
  \BibitemOpen
  \bibfield  {author} {\bibinfo {author} {\bibfnamefont {Y.}~\bibnamefont
  {Tserkovnyak}}, \bibinfo {author} {\bibfnamefont {A.}~\bibnamefont
  {Brataas}}, \ and\ \bibinfo {author} {\bibfnamefont {G.~E.~W.}\ \bibnamefont
  {Bauer}},\ }\href {\doibase 10.1103/PhysRevLett.88.117601} {\bibfield
  {journal} {\bibinfo  {journal} {Phys. Rev. Lett.}\ }\textbf {\bibinfo
  {volume} {88}},\ \bibinfo {pages} {117601} (\bibinfo {year}
  {2002}{\natexlab{b}})}\BibitemShut {NoStop}%
\bibitem [{\citenamefont {Tserkovnyak}\ \emph {et~al.}(2003)\citenamefont
  {Tserkovnyak}, \citenamefont {Brataas},\ and\ \citenamefont
  {Bauer}}]{tserkovnyakDynamicExchangeCoupling2003}%
  \BibitemOpen
  \bibfield  {author} {\bibinfo {author} {\bibfnamefont {Y.}~\bibnamefont
  {Tserkovnyak}}, \bibinfo {author} {\bibfnamefont {A.}~\bibnamefont
  {Brataas}}, \ and\ \bibinfo {author} {\bibfnamefont {G.~E.~W.}\ \bibnamefont
  {Bauer}},\ }\href {\doibase 10.1063/1.1538173} {\bibfield  {journal}
  {\bibinfo  {journal} {Journal of Applied Physics}\ }\textbf {\bibinfo
  {volume} {93}},\ \bibinfo {pages} {7534} (\bibinfo {year}
  {2003})}\BibitemShut {NoStop}%
\bibitem [{\citenamefont {Mucciolo}\ \emph {et~al.}(2002)\citenamefont
  {Mucciolo}, \citenamefont {Chamon},\ and\ \citenamefont
  {Marcus}}]{muccioloAdiabaticQuantumPump2002}%
  \BibitemOpen
  \bibfield  {author} {\bibinfo {author} {\bibfnamefont {E.~R.}\ \bibnamefont
  {Mucciolo}}, \bibinfo {author} {\bibfnamefont {C.}~\bibnamefont {Chamon}}, \
  and\ \bibinfo {author} {\bibfnamefont {C.~M.}\ \bibnamefont {Marcus}},\
  }\href {\doibase 10.1103/PhysRevLett.89.146802} {\bibfield  {journal}
  {\bibinfo  {journal} {Phys. Rev. Lett.}\ }\textbf {\bibinfo {volume} {89}},\
  \bibinfo {pages} {146802} (\bibinfo {year} {2002})}\BibitemShut {NoStop}%
\bibitem [{\citenamefont {Sharma}\ and\ \citenamefont
  {Chamon}(2003)}]{sharmaAdiabaticChargeSpin2003}%
  \BibitemOpen
  \bibfield  {author} {\bibinfo {author} {\bibfnamefont {P.}~\bibnamefont
  {Sharma}}\ and\ \bibinfo {author} {\bibfnamefont {C.}~\bibnamefont
  {Chamon}},\ }\href {\doibase 10.1103/PhysRevB.68.035321} {\bibfield
  {journal} {\bibinfo  {journal} {Phys. Rev. B}\ }\textbf {\bibinfo {volume}
  {68}},\ \bibinfo {pages} {035321} (\bibinfo {year} {2003})}\BibitemShut
  {NoStop}%
\bibitem [{\citenamefont {Watson}\ \emph {et~al.}(2003)\citenamefont {Watson},
  \citenamefont {Potok}, \citenamefont {Marcus},\ and\ \citenamefont
  {Umansky}}]{watsonExperimentalRealizationQuantum2003}%
  \BibitemOpen
  \bibfield  {author} {\bibinfo {author} {\bibfnamefont {S.~K.}\ \bibnamefont
  {Watson}}, \bibinfo {author} {\bibfnamefont {R.~M.}\ \bibnamefont {Potok}},
  \bibinfo {author} {\bibfnamefont {C.~M.}\ \bibnamefont {Marcus}}, \ and\
  \bibinfo {author} {\bibfnamefont {V.}~\bibnamefont {Umansky}},\ }\href
  {\doibase 10.1103/PhysRevLett.91.258301} {\bibfield  {journal} {\bibinfo
  {journal} {Phys. Rev. Lett.}\ }\textbf {\bibinfo {volume} {91}},\ \bibinfo
  {pages} {258301} (\bibinfo {year} {2003})}\BibitemShut {NoStop}%
\bibitem [{\citenamefont {Xia}\ \emph {et~al.}(2002)\citenamefont {Xia},
  \citenamefont {Kelly}, \citenamefont {Bauer}, \citenamefont {Brataas},\ and\
  \citenamefont {Turek}}]{xiaSpinTorquesFerromagnetic2002}%
  \BibitemOpen
  \bibfield  {author} {\bibinfo {author} {\bibfnamefont {K.}~\bibnamefont
  {Xia}}, \bibinfo {author} {\bibfnamefont {P.~J.}\ \bibnamefont {Kelly}},
  \bibinfo {author} {\bibfnamefont {G.~E.~W.}\ \bibnamefont {Bauer}}, \bibinfo
  {author} {\bibfnamefont {A.}~\bibnamefont {Brataas}}, \ and\ \bibinfo
  {author} {\bibfnamefont {I.}~\bibnamefont {Turek}},\ }\href {\doibase
  10.1103/PhysRevB.65.220401} {\bibfield  {journal} {\bibinfo  {journal} {Phys.
  Rev. B}\ }\textbf {\bibinfo {volume} {65}},\ \bibinfo {pages} {220401}
  (\bibinfo {year} {2002})}\BibitemShut {NoStop}%
\bibitem [{\citenamefont {Ohnuma}\ \emph {et~al.}(2014)\citenamefont {Ohnuma},
  \citenamefont {Adachi}, \citenamefont {Saitoh},\ and\ \citenamefont
  {Maekawa}}]{ohnumaEnhanced2014}%
  \BibitemOpen
  \bibfield  {author} {\bibinfo {author} {\bibfnamefont {Y.}~\bibnamefont
  {Ohnuma}}, \bibinfo {author} {\bibfnamefont {H.}~\bibnamefont {Adachi}},
  \bibinfo {author} {\bibfnamefont {E.}~\bibnamefont {Saitoh}}, \ and\ \bibinfo
  {author} {\bibfnamefont {S.}~\bibnamefont {Maekawa}},\ }\href {\doibase
  10.1103/PhysRevB.89.174417} {\bibfield  {journal} {\bibinfo  {journal} {Phys.
  Rev. B}\ }\textbf {\bibinfo {volume} {89}},\ \bibinfo {pages} {174417}
  (\bibinfo {year} {2014})}\BibitemShut {NoStop}%
\bibitem [{\citenamefont {Matsuo}\ \emph {et~al.}(2018)\citenamefont {Matsuo},
  \citenamefont {Ohnuma}, \citenamefont {Kato},\ and\ \citenamefont
  {Maekawa}}]{matsuo2018Spin}%
  \BibitemOpen
  \bibfield  {author} {\bibinfo {author} {\bibfnamefont {M.}~\bibnamefont
  {Matsuo}}, \bibinfo {author} {\bibfnamefont {Y.}~\bibnamefont {Ohnuma}},
  \bibinfo {author} {\bibfnamefont {T.}~\bibnamefont {Kato}}, \ and\ \bibinfo
  {author} {\bibfnamefont {S.}~\bibnamefont {Maekawa}},\ }\href {\doibase
  10.1103/PhysRevLett.120.037201} {\bibfield  {journal} {\bibinfo  {journal}
  {Phys. Rev. Lett.}\ }\textbf {\bibinfo {volume} {120}},\ \bibinfo {pages}
  {037201} (\bibinfo {year} {2018})}\BibitemShut {NoStop}%
\bibitem [{\citenamefont {Kato}\ \emph {et~al.}(2019)\citenamefont {Kato},
  \citenamefont {Ohnuma}, \citenamefont {Matsuo}, \citenamefont {Rech},
  \citenamefont {Jonckheere},\ and\ \citenamefont
  {Martin}}]{katoMicroscopicTheorySpin2019}%
  \BibitemOpen
  \bibfield  {author} {\bibinfo {author} {\bibfnamefont {T.}~\bibnamefont
  {Kato}}, \bibinfo {author} {\bibfnamefont {Y.}~\bibnamefont {Ohnuma}},
  \bibinfo {author} {\bibfnamefont {M.}~\bibnamefont {Matsuo}}, \bibinfo
  {author} {\bibfnamefont {J.}~\bibnamefont {Rech}}, \bibinfo {author}
  {\bibfnamefont {T.}~\bibnamefont {Jonckheere}}, \ and\ \bibinfo {author}
  {\bibfnamefont {T.}~\bibnamefont {Martin}},\ }\href {\doibase
  10.1103/PhysRevB.99.144411} {\bibfield  {journal} {\bibinfo  {journal} {Phys.
  Rev. B}\ }\textbf {\bibinfo {volume} {99}},\ \bibinfo {pages} {144411}
  (\bibinfo {year} {2019})}\BibitemShut {NoStop}%
\bibitem [{\citenamefont {Kato}\ \emph {et~al.}(2020)\citenamefont {Kato},
  \citenamefont {Ohnuma},\ and\ \citenamefont
  {Matsuo}}]{katoMicroscopicTheorySpin2020}%
  \BibitemOpen
  \bibfield  {author} {\bibinfo {author} {\bibfnamefont {T.}~\bibnamefont
  {Kato}}, \bibinfo {author} {\bibfnamefont {Y.}~\bibnamefont {Ohnuma}}, \ and\
  \bibinfo {author} {\bibfnamefont {M.}~\bibnamefont {Matsuo}},\ }\href
  {\doibase 10.1103/PhysRevB.102.094437} {\bibfield  {journal} {\bibinfo
  {journal} {Phys. Rev. B}\ }\textbf {\bibinfo {volume} {102}},\ \bibinfo
  {pages} {094437} (\bibinfo {year} {2020})}\BibitemShut {NoStop}%
\bibitem [{\citenamefont {Ominato}\ and\ \citenamefont
  {Matsuo}(2020)}]{ominatoQuantumOscillationsGilbert2020}%
  \BibitemOpen
  \bibfield  {author} {\bibinfo {author} {\bibfnamefont {Y.}~\bibnamefont
  {Ominato}}\ and\ \bibinfo {author} {\bibfnamefont {M.}~\bibnamefont
  {Matsuo}},\ }\href {\doibase 10.7566/JPSJ.89.053704} {\bibfield  {journal}
  {\bibinfo  {journal} {J. Phys. Soc. Jpn.}\ }\textbf {\bibinfo {volume}
  {89}},\ \bibinfo {pages} {053704} (\bibinfo {year} {2020})}\BibitemShut
  {NoStop}%
\bibitem [{\citenamefont {Ominato}\ \emph {et~al.}(2020)\citenamefont
  {Ominato}, \citenamefont {Fujimoto},\ and\ \citenamefont
  {Matsuo}}]{ominatoValleyDependentSpinTransport2020}%
  \BibitemOpen
  \bibfield  {author} {\bibinfo {author} {\bibfnamefont {Y.}~\bibnamefont
  {Ominato}}, \bibinfo {author} {\bibfnamefont {J.}~\bibnamefont {Fujimoto}}, \
  and\ \bibinfo {author} {\bibfnamefont {M.}~\bibnamefont {Matsuo}},\ }\href
  {\doibase 10.1103/PhysRevLett.124.166803} {\bibfield  {journal} {\bibinfo
  {journal} {Phys. Rev. Lett.}\ }\textbf {\bibinfo {volume} {124}},\ \bibinfo
  {pages} {166803} (\bibinfo {year} {2020})}\BibitemShut {NoStop}%
\bibitem [{\citenamefont {Yamamoto}\ \emph {et~al.}(2021)\citenamefont
  {Yamamoto}, \citenamefont {Kato},\ and\ \citenamefont
  {Matsuo}}]{yamamotoSpinCurrentMagnetic2021}%
  \BibitemOpen
  \bibfield  {author} {\bibinfo {author} {\bibfnamefont {T.}~\bibnamefont
  {Yamamoto}}, \bibinfo {author} {\bibfnamefont {T.}~\bibnamefont {Kato}}, \
  and\ \bibinfo {author} {\bibfnamefont {M.}~\bibnamefont {Matsuo}},\ }\href
  {\doibase 10.1103/PhysRevB.104.L121401} {\bibfield  {journal} {\bibinfo
  {journal} {Phys. Rev. B}\ }\textbf {\bibinfo {volume} {104}},\ \bibinfo
  {pages} {L121401} (\bibinfo {year} {2021})},\ \Eprint
  {http://arxiv.org/abs/2106.06102} {arXiv:2106.06102} \BibitemShut {NoStop}%
\bibitem{ominatoAnisotropicSuperconductingSpin2021}
Y. Ominato, A. Yamakage, and M. Matsuo, arXiv:2103.05871 [cond-mat] (2021).
\bibitem [{\citenamefont {Yama}\ \emph {et~al.}(2021)\citenamefont {Yama},
  \citenamefont {Tatsuno}, \citenamefont {Kato},\ and\ \citenamefont
  {Matsuo}}]{yama2021Spin}%
  \BibitemOpen
  \bibfield  {author} {\bibinfo {author} {\bibfnamefont {M.}~\bibnamefont
  {Yama}}, \bibinfo {author} {\bibfnamefont {M.}~\bibnamefont {Tatsuno}},
  \bibinfo {author} {\bibfnamefont {T.}~\bibnamefont {Kato}}, \ and\ \bibinfo
  {author} {\bibfnamefont {M.}~\bibnamefont {Matsuo}},\ }\href {\doibase
  10.1103/PhysRevB.104.054410} {\bibfield  {journal} {\bibinfo  {journal}
  {Phys. Rev. B}\ }\textbf {\bibinfo {volume} {104}},\ \bibinfo {pages}
  {054410} (\bibinfo {year} {2021})}\BibitemShut {NoStop}%
\bibitem{Yama2022} M. Yama, M. Matsuo, and T. Kato, arXiv:2201.11498 [cond-mat] (2022)
\bibitem [{\citenamefont {Ominato}\ \emph {et~al.}(2022)\citenamefont
  {Ominato}, \citenamefont {Yamakage}, \citenamefont {Kato},\ and\
  \citenamefont {Matsuo}}]{ominato2022Ferromagnetica}%
  \BibitemOpen
  \bibfield  {author} {\bibinfo {author} {\bibfnamefont {Y.}~\bibnamefont
  {Ominato}}, \bibinfo {author} {\bibfnamefont {A.}~\bibnamefont {Yamakage}},
  \bibinfo {author} {\bibfnamefont {T.}~\bibnamefont {Kato}}, \ and\ \bibinfo
  {author} {\bibfnamefont {M.}~\bibnamefont {Matsuo}},\ }\href {\doibase
  10.1103/PhysRevB.105.205406} {\bibfield  {journal} {\bibinfo  {journal}
  {Phys. Rev. B}\ }\textbf {\bibinfo {volume} {105}},\ \bibinfo {pages}
  {205406} (\bibinfo {year} {2022})}\BibitemShut {NoStop}%
\bibitem [{\citenamefont {{\'E}del'man}(1976)}]{edelmanElectronsBismuth1976}%
  \BibitemOpen
  \bibfield  {author} {\bibinfo {author} {\bibfnamefont {V.}~\bibnamefont
  {{\'E}del'man}},\ }\href {\doibase 10.1080/00018737600101452} {\bibfield
  {journal} {\bibinfo  {journal} {Advances in Physics}\ }\textbf {\bibinfo
  {volume} {25}},\ \bibinfo {pages} {555} (\bibinfo {year} {1976})}\BibitemShut
  {NoStop}%
\bibitem [{\citenamefont {Fuseya}\ \emph {et~al.}(2015)\citenamefont {Fuseya},
  \citenamefont {Ogata},\ and\ \citenamefont
  {Fukuyama}}]{fuseyaTransportPropertiesDiamagnetism2015}%
  \BibitemOpen
  \bibfield  {author} {\bibinfo {author} {\bibfnamefont {Y.}~\bibnamefont
  {Fuseya}}, \bibinfo {author} {\bibfnamefont {M.}~\bibnamefont {Ogata}}, \
  and\ \bibinfo {author} {\bibfnamefont {H.}~\bibnamefont {Fukuyama}},\ }\href
  {\doibase 10.7566/JPSJ.84.012001} {\bibfield  {journal} {\bibinfo  {journal}
  {J. Phys. Soc. Jpn.}\ }\textbf {\bibinfo {volume} {84}},\ \bibinfo {pages}
  {012001} (\bibinfo {year} {2015})}\BibitemShut {NoStop}%
\bibitem [{\citenamefont {Wolff}(1964)}]{wolffMatrixElementsSelection1964}%
  \BibitemOpen
  \bibfield  {author} {\bibinfo {author} {\bibfnamefont {P.~A.}\ \bibnamefont
  {Wolff}},\ }\href {\doibase 10.1016/0022-3697(64)90128-3} {\bibfield
  {journal} {\bibinfo  {journal} {Journal of Physics and Chemistry of Solids}\
  }\textbf {\bibinfo {volume} {25}},\ \bibinfo {pages} {1057} (\bibinfo {year}
  {1964})}\BibitemShut {NoStop}%
\bibitem [{\citenamefont {Fu}\ and\ \citenamefont
  {Kane}(2007)}]{fuTopologicalInsulatorsInversion2007}%
  \BibitemOpen
  \bibfield  {author} {\bibinfo {author} {\bibfnamefont {L.}~\bibnamefont
  {Fu}}\ and\ \bibinfo {author} {\bibfnamefont {C.~L.}\ \bibnamefont {Kane}},\
  }\href {\doibase 10.1103/PhysRevB.76.045302} {\bibfield  {journal} {\bibinfo
  {journal} {Phys. Rev. B}\ }\textbf {\bibinfo {volume} {76}},\ \bibinfo
  {pages} {045302} (\bibinfo {year} {2007})}\BibitemShut {NoStop}%
\bibitem [{\citenamefont {Teo}\ \emph {et~al.}(2008)\citenamefont {Teo},
  \citenamefont {Fu},\ and\ \citenamefont
  {Kane}}]{teoSurfaceStatesTopological2008}%
  \BibitemOpen
  \bibfield  {author} {\bibinfo {author} {\bibfnamefont {J.~C.~Y.}\
  \bibnamefont {Teo}}, \bibinfo {author} {\bibfnamefont {L.}~\bibnamefont
  {Fu}}, \ and\ \bibinfo {author} {\bibfnamefont {C.~L.}\ \bibnamefont
  {Kane}},\ }\href {\doibase 10.1103/PhysRevB.78.045426} {\bibfield  {journal}
  {\bibinfo  {journal} {Phys. Rev. B}\ }\textbf {\bibinfo {volume} {78}},\
  \bibinfo {pages} {045426} (\bibinfo {year} {2008})}\BibitemShut {NoStop}%
\bibitem [{\citenamefont {Fuseya}\ \emph
  {et~al.}(2012{\natexlab{a}})\citenamefont {Fuseya}, \citenamefont {Ogata},\
  and\ \citenamefont {Fukuyama}}]{fuseyaSpinHallEffectDiamagnetism2012}%
  \BibitemOpen
  \bibfield  {author} {\bibinfo {author} {\bibfnamefont {Y.}~\bibnamefont
  {Fuseya}}, \bibinfo {author} {\bibfnamefont {M.}~\bibnamefont {Ogata}}, \
  and\ \bibinfo {author} {\bibfnamefont {H.}~\bibnamefont {Fukuyama}},\ }\href
  {\doibase 10.1143/JPSJ.81.093704} {\bibfield  {journal} {\bibinfo  {journal}
  {J. Phys. Soc. Jpn.}\ }\textbf {\bibinfo {volume} {81}},\ \bibinfo {pages}
  {093704} (\bibinfo {year} {2012}{\natexlab{a}})}\BibitemShut {NoStop}%
\bibitem [{\citenamefont {Fuseya}\ \emph
  {et~al.}(2012{\natexlab{b}})\citenamefont {Fuseya}, \citenamefont {Ogata},\
  and\ \citenamefont
  {Fukuyama}}]{fuseyaSpinPolarizationMagnetoOpticalConductivity2012}%
  \BibitemOpen
  \bibfield  {author} {\bibinfo {author} {\bibfnamefont {Y.}~\bibnamefont
  {Fuseya}}, \bibinfo {author} {\bibfnamefont {M.}~\bibnamefont {Ogata}}, \
  and\ \bibinfo {author} {\bibfnamefont {H.}~\bibnamefont {Fukuyama}},\ }\href
  {\doibase 10.1143/JPSJ.81.013704} {\bibfield  {journal} {\bibinfo  {journal}
  {J. Phys. Soc. Jpn.}\ }\textbf {\bibinfo {volume} {81}},\ \bibinfo {pages}
  {013704} (\bibinfo {year} {2012}{\natexlab{b}})}\BibitemShut {NoStop}%
\bibitem [{\citenamefont {Fuseya}\ \emph {et~al.}(2014)\citenamefont {Fuseya},
  \citenamefont {Ogata},\ and\ \citenamefont
  {Fukuyama}}]{fuseyaSpinHallEffectDiamagnetism2014}%
  \BibitemOpen
  \bibfield  {author} {\bibinfo {author} {\bibfnamefont {Y.}~\bibnamefont
  {Fuseya}}, \bibinfo {author} {\bibfnamefont {M.}~\bibnamefont {Ogata}}, \
  and\ \bibinfo {author} {\bibfnamefont {H.}~\bibnamefont {Fukuyama}},\ }\href
  {\doibase 10.7566/JPSJ.83.074702} {\bibfield  {journal} {\bibinfo  {journal}
  {J. Phys. Soc. Jpn.}\ }\textbf {\bibinfo {volume} {83}},\ \bibinfo {pages}
  {074702} (\bibinfo {year} {2014})}\BibitemShut {NoStop}%
\bibitem [{\citenamefont {Fukazawa}\ \emph {et~al.}(2017)\citenamefont
  {Fukazawa}, \citenamefont {Kohno},\ and\ \citenamefont
  {Fujimoto}}]{fukazawaIntrinsicExtrinsicSpin2017}%
  \BibitemOpen
  \bibfield  {author} {\bibinfo {author} {\bibfnamefont {T.}~\bibnamefont
  {Fukazawa}}, \bibinfo {author} {\bibfnamefont {H.}~\bibnamefont {Kohno}}, \
  and\ \bibinfo {author} {\bibfnamefont {J.}~\bibnamefont {Fujimoto}},\ }\href
  {\doibase 10.7566/JPSJ.86.094704} {\bibfield  {journal} {\bibinfo  {journal}
  {J. Phys. Soc. Jpn.}\ }\textbf {\bibinfo {volume} {86}},\ \bibinfo {pages}
  {094704} (\bibinfo {year} {2017})}\BibitemShut {NoStop}%
\bibitem [{\citenamefont {Yue}\ \emph {et~al.}(2018)\citenamefont {Yue},
  \citenamefont {Lin}, \citenamefont {Li}, \citenamefont {Jin},\ and\
  \citenamefont {Chien}}]{yueSpintoChargeConversionBi2018}%
  \BibitemOpen
  \bibfield  {author} {\bibinfo {author} {\bibfnamefont {D.}~\bibnamefont
  {Yue}}, \bibinfo {author} {\bibfnamefont {W.}~\bibnamefont {Lin}}, \bibinfo
  {author} {\bibfnamefont {J.}~\bibnamefont {Li}}, \bibinfo {author}
  {\bibfnamefont {X.}~\bibnamefont {Jin}}, \ and\ \bibinfo {author}
  {\bibfnamefont {C.~L.}\ \bibnamefont {Chien}},\ }\href {\doibase
  10.1103/PhysRevLett.121.037201} {\bibfield  {journal} {\bibinfo  {journal}
  {Phys. Rev. Lett.}\ }\textbf {\bibinfo {volume} {121}},\ \bibinfo {pages}
  {037201} (\bibinfo {year} {2018})}\BibitemShut {NoStop}%
\bibitem [{\citenamefont {Chi}\ \emph {et~al.}(2020)\citenamefont {Chi},
  \citenamefont {Lau}, \citenamefont {Xu}, \citenamefont {Ohkubo},
  \citenamefont {Hono},\ and\ \citenamefont {Hayashi}}]{chiSpinHallEffect2020}%
  \BibitemOpen
  \bibfield  {author} {\bibinfo {author} {\bibfnamefont {Z.}~\bibnamefont
  {Chi}}, \bibinfo {author} {\bibfnamefont {Y.-C.}\ \bibnamefont {Lau}},
  \bibinfo {author} {\bibfnamefont {X.}~\bibnamefont {Xu}}, \bibinfo {author}
  {\bibfnamefont {T.}~\bibnamefont {Ohkubo}}, \bibinfo {author} {\bibfnamefont
  {K.}~\bibnamefont {Hono}}, \ and\ \bibinfo {author} {\bibfnamefont
  {M.}~\bibnamefont {Hayashi}},\ }\href {\doibase 10.1126/sciadv.aay2324}
  {\bibfield  {journal} {\bibinfo  {journal} {Sci. Adv.}\ }\textbf {\bibinfo
  {volume} {6}},\ \bibinfo {pages} {eaay2324} (\bibinfo {year}
  {2020})}\BibitemShut {NoStop}%
\bibitem [{\citenamefont {Karube}\ \emph {et~al.}(2016)\citenamefont {Karube},
  \citenamefont {Kondou},\ and\ \citenamefont
  {Otani}}]{karubeExperimentalObservationSpintocharge2016}%
  \BibitemOpen
  \bibfield  {author} {\bibinfo {author} {\bibfnamefont {S.}~\bibnamefont
  {Karube}}, \bibinfo {author} {\bibfnamefont {K.}~\bibnamefont {Kondou}}, \
  and\ \bibinfo {author} {\bibfnamefont {Y.}~\bibnamefont {Otani}},\ }\href
  {\doibase 10.7567/APEX.9.033001} {\bibfield  {journal} {\bibinfo  {journal}
  {Appl. Phys. Express}\ }\textbf {\bibinfo {volume} {9}},\ \bibinfo {pages}
  {033001} (\bibinfo {year} {2016})}\BibitemShut {NoStop}%
\bibitem [{\citenamefont {Hou}\ \emph {et~al.}(2012)\citenamefont {Hou},
  \citenamefont {Qiu}, \citenamefont {Harii}, \citenamefont {Kajiwara},
  \citenamefont {Uchida}, \citenamefont {Fujikawa}, \citenamefont {Nakayama},
  \citenamefont {Yoshino}, \citenamefont {An}, \citenamefont {Ando},
  \citenamefont {Jin},\ and\ \citenamefont
  {Saitoh}}]{houInterfaceInducedInverse2012}%
  \BibitemOpen
  \bibfield  {author} {\bibinfo {author} {\bibfnamefont {D.}~\bibnamefont
  {Hou}}, \bibinfo {author} {\bibfnamefont {Z.}~\bibnamefont {Qiu}}, \bibinfo
  {author} {\bibfnamefont {K.}~\bibnamefont {Harii}}, \bibinfo {author}
  {\bibfnamefont {Y.}~\bibnamefont {Kajiwara}}, \bibinfo {author}
  {\bibfnamefont {K.}~\bibnamefont {Uchida}}, \bibinfo {author} {\bibfnamefont
  {Y.}~\bibnamefont {Fujikawa}}, \bibinfo {author} {\bibfnamefont
  {H.}~\bibnamefont {Nakayama}}, \bibinfo {author} {\bibfnamefont
  {T.}~\bibnamefont {Yoshino}}, \bibinfo {author} {\bibfnamefont
  {T.}~\bibnamefont {An}}, \bibinfo {author} {\bibfnamefont {K.}~\bibnamefont
  {Ando}}, \bibinfo {author} {\bibfnamefont {X.}~\bibnamefont {Jin}}, \ and\
  \bibinfo {author} {\bibfnamefont {E.}~\bibnamefont {Saitoh}},\ }\href
  {\doibase 10.1063/1.4738786} {\bibfield  {journal} {\bibinfo  {journal}
  {Appl. Phys. Lett.}\ }\textbf {\bibinfo {volume} {101}},\ \bibinfo {pages}
  {042403} (\bibinfo {year} {2012})}\BibitemShut {NoStop}%
\bibitem [{\citenamefont {Emoto}\ \emph {et~al.}(2014)\citenamefont {Emoto},
  \citenamefont {Ando}, \citenamefont {Shikoh}, \citenamefont {Fuseya},
  \citenamefont {Shinjo},\ and\ \citenamefont
  {Shiraishi}}]{emotoConversionPureSpin2014}%
  \BibitemOpen
  \bibfield  {author} {\bibinfo {author} {\bibfnamefont {H.}~\bibnamefont
  {Emoto}}, \bibinfo {author} {\bibfnamefont {Y.}~\bibnamefont {Ando}},
  \bibinfo {author} {\bibfnamefont {E.}~\bibnamefont {Shikoh}}, \bibinfo
  {author} {\bibfnamefont {Y.}~\bibnamefont {Fuseya}}, \bibinfo {author}
  {\bibfnamefont {T.}~\bibnamefont {Shinjo}}, \ and\ \bibinfo {author}
  {\bibfnamefont {M.}~\bibnamefont {Shiraishi}},\ }\href {\doibase
  10.1063/1.4863377} {\bibfield  {journal} {\bibinfo  {journal} {Journal of
  Applied Physics}\ }\textbf {\bibinfo {volume} {115}},\ \bibinfo {pages}
  {17C507} (\bibinfo {year} {2014})}\BibitemShut {NoStop}%
\bibitem [{\citenamefont {Emoto}\ \emph {et~al.}(2016)\citenamefont {Emoto},
  \citenamefont {Ando}, \citenamefont {Eguchi}, \citenamefont {Ohshima},
  \citenamefont {Shikoh}, \citenamefont {Fuseya}, \citenamefont {Shinjo},\ and\
  \citenamefont {Shiraishi}}]{emotoTransportSpinConversion2016}%
  \BibitemOpen
  \bibfield  {author} {\bibinfo {author} {\bibfnamefont {H.}~\bibnamefont
  {Emoto}}, \bibinfo {author} {\bibfnamefont {Y.}~\bibnamefont {Ando}},
  \bibinfo {author} {\bibfnamefont {G.}~\bibnamefont {Eguchi}}, \bibinfo
  {author} {\bibfnamefont {R.}~\bibnamefont {Ohshima}}, \bibinfo {author}
  {\bibfnamefont {E.}~\bibnamefont {Shikoh}}, \bibinfo {author} {\bibfnamefont
  {Y.}~\bibnamefont {Fuseya}}, \bibinfo {author} {\bibfnamefont
  {T.}~\bibnamefont {Shinjo}}, \ and\ \bibinfo {author} {\bibfnamefont
  {M.}~\bibnamefont {Shiraishi}},\ }\href {\doibase 10.1103/PhysRevB.93.174428}
  {\bibfield  {journal} {\bibinfo  {journal} {Phys. Rev. B}\ }\textbf {\bibinfo
  {volume} {93}},\ \bibinfo {pages} {174428} (\bibinfo {year}
  {2016})}\BibitemShut {NoStop}%
\bibitem [{\citenamefont {Tang}\ and\ \citenamefont
  {S.~Dresselhaus}(2014)}]{tangElectronicPropertiesNanostructured2014}%
  \BibitemOpen
  \bibfield  {author} {\bibinfo {author} {\bibfnamefont {S.}~\bibnamefont
  {Tang}}\ and\ \bibinfo {author} {\bibfnamefont {M.}~\bibnamefont
  {S.~Dresselhaus}},\ }\href {\doibase 10.1039/C4TC00146J} {\bibfield
  {journal} {\bibinfo  {journal} {Journal of Materials Chemistry C}\ }\textbf
  {\bibinfo {volume} {2}},\ \bibinfo {pages} {4710} (\bibinfo {year}
  {2014})}\BibitemShut {NoStop}%
\bibitem [{\citenamefont {Zhu}\ \emph {et~al.}(2011)\citenamefont {Zhu},
  \citenamefont {Fauqu{\'e}}, \citenamefont {Fuseya},\ and\ \citenamefont
  {Behnia}}]{zhu2011Angleresolved}%
  \BibitemOpen
  \bibfield  {author} {\bibinfo {author} {\bibfnamefont {Z.}~\bibnamefont
  {Zhu}}, \bibinfo {author} {\bibfnamefont {B.}~\bibnamefont {Fauqu{\'e}}},
  \bibinfo {author} {\bibfnamefont {Y.}~\bibnamefont {Fuseya}}, \ and\ \bibinfo
  {author} {\bibfnamefont {K.}~\bibnamefont {Behnia}},\ }\href {\doibase
  10.1103/PhysRevB.84.115137} {\bibfield  {journal} {\bibinfo  {journal} {Phys.
  Rev. B}\ }\textbf {\bibinfo {volume} {84}},\ \bibinfo {pages} {115137}
  (\bibinfo {year} {2011})}\BibitemShut {NoStop}%
\end{thebibliography}

%

\end{document}